\documentclass[nonacm,screen]{acmart}
\pdfoutput=1
\usepackage{graphicx}
\usepackage{pifont}
\usepackage{multirow}
\usepackage{fontawesome5}

\newcommand\ImageCheck{$\vcenter{\hbox{\scriptsize\faImage}}$}
\newcommand\VideoCheck{\;$\vcenter{\hbox{\scriptsize\faVideo}}$}

\newcommand\same{\multicolumn{1}{c}{$\equiv$}}

\newcommand\qtiny{\epsilon}

\definecolor{lowlight}{gray}{0.55}
\definecolor{slate}{HTML}{864b79}
\definecolor{DeepRed}{HTML}{8a0312}
\definecolor{DeepGreen}{HTML}{0a580f}

\makeatletter
\renewcommand{\anon}[2][ANONYMIZED]{\if@ACM@anonymous #1\else #2\fi}
\makeatother

\newcommand\hidden[1]{}
\newcommand\V[1]{\textsc{\MakeLowercase{#1}}}

\newcommand\yes{\ding{52}}

\theoremstyle{definition}
\newtheorem{finding}{Finding}

\newif\iftechreport
\techreporttrue
\newcommand\TechReport[1]{\iftechreport {#1}\fi}
\newcommand\Journal[1]{\iftechreport\else{#1}\fi}

\AtBeginDocument{\libertineOsF}

\begin{document}

\title{Putting the Count Back Into Accountability}
\subtitle{An Analysis of Transparency Data About the Sexual Exploitation of Minors}

\author{Robert Grimm}
\orcid{0000-0002-8300-2153}
\affiliation{\institution{Charles University}
    \city{Prague}
    \country{Czech Republic}
}
\email{rgrimm@alum.mit.edu}

\begin{abstract}

Alarmist and sensationalist statements about the ``explosion'' of online child
sexual exploitation or \V{CSE} dominate much of the public discourse about the
topic. Based on a new dataset collecting the transparency disclosures for 16
\V{US}-based internet platforms and the national clearinghouse collecting
legally mandated reports about \V{CSE}, this study seeks answers to two research
questions: First, what does the data tell us about the growth of online \V{CSE}?
Second, how reliable and trustworthy is that data? To answer the two questions,
this study proceeds in three parts. First, we leverage a critical literature
review to synthesize a granular model for \V{CSE} reporting. Second, we analyze
the growth in \V{CSE} reports over the last 25~years and correlate it with the
growth of social media user accounts. Third, we use two comparative audits to
assess the quality of transparency data. Critical findings include: First,
\V{US} law increasingly threatens the very population it claims to protect,
i.e., children and adolescents. Second, the rapid growth of \V{CSE} report over
the last decade is linear and largely driven by an equivalent growth in social
media user accounts. Third, the Covid-19 pandemic had no statistically relevant
impact on report volume. Fourth, while half of surveyed organizations release
meaningful and reasonably accurate transparency data, the other half either fail
to make disclosures or release data with severe quality issues.

\end{abstract}


\maketitle

\section{Introduction}
\label{sec:introduction}

Public debate about child sexual exploitation (\V{CSE}) features many adamant to
alarming claims. There are the headlines in national newspapers announcing that
``the internet is overrun with images of child sexual
abuse''~\cite{KellerDance2019}, ``Instagram connects vast pedophile
network''~\cite{HorwitzBlunt2023}, and ``\V{AI} is about to make the online
child sex abuse problem much worse''~\cite{Oremus2024}. There are Senate
hearings on ``big tech and the online child sexual exploitation crisis''
featuring a who-is-who of social media \V{CEO}s~\cite{Durbin2024} and laws
referencing the ``explosion in the multijurisdictional distribution of child
pornography''~\cite{UnitedStatesCongress2008}. There also are the Moms for
Liberty, Qanon adherents, and Proud Boys mobilizing to ``\#SaveTheChildren''
while maligning everyone else, particularly \V{LGBT} folk, as ``pedos'' or
``groomers''~\cite{BuntainBarlowea2022, MartinyLawrence2023}. By contrast,
experts such as academics~\cite{SalterWoodlockea2023},
lawyers~\cite{Haney2021b}, hotline operators~\cite{ODonnell2021}, and
police~\cite{Europol2020, Interpol2020} tend to blame the Covid-19 pandemic,
often in unusually lurid terms.

Given the gravity of the topic and the creeping sensationalism infusing even the
law, two questions stand out:
\begin{enumerate}
\item What does available data tell us about the extent of online child sexual
exploitation?
\item How accurate and trustworthy is that data?
\end{enumerate}
This article presents the results of a study that seeks to answer these two
research questions based on the transparency disclosures (or lack thereof) of
15~technology firms, including popular social media platforms, and
2~not-for-profits, including the national clearinghouse for reports about
\V{CSE}, the National Center for Missing and Exploited Children or \V{NCMEC}.
According to \V{US} law, service providers needn't proactively scan for child
sexual abuse materials (\V{CSAM}) and other exploitative activities. But when
they become aware of them, they must file so-called CyberTipline reports in an
expedient manner.

The study comprises three parts, one to establish context and one each to answer
the two research questions. In particular, the first part is based on a critical
literature review that draws on a broad spectrum of sources not limited to the
academic literature. The primary goals were to synthesize a conceptual model of
CyberTipline reporting and to identify gaps in our knowledge. The second part is
an analysis of the yearly CyberTipline report volume over the last quarter
century. That includes determining a suitable regression model and identifying
the likely primary force driving the rapid growth over the last decade. The
third part features two comparative audits of transparency data. Both audits
leverage redundant or repeated disclosures of what should be the same quantities
to assess data quality. At the same time, they differ significantly in what
constitutes (un)acceptable divergence.

Significant findings include:
\begin{enumerate}
\item A significant overreach of \V{US} law that increasingly threatens the very
population it claims to protect, i.e., children and adolescents;
\item Linear growth in CyberTipline reports over the last decade that is largely
driven by a corresponding increase in social media user accounts;
\item No statistically relevant impact of the pandemic on report volume;
\item A half-half split between firms that release meaningful and reasonably
accurate transparency data and those that either fail to make disclosures or
release data with severe quality issues.
\end{enumerate}
The second finding directly contradicts claims of an ``explosion'' of \V{CSAM}
or an internet ``overrun with images of child sexual abuse.'' But that doesn't
change that the deluge of reports is bringing \V{NCMEC} close to breaking
point~\cite{GrossmanPfefferkornea2024}. As discussed in
Section~\ref{sec:discussion}, the findings of this study not only allow us to
provide a bound on possible future growth. But they also suggest two approaches
for reigning in the deluge of CyberTipline reports. Alas, evaluating their
impact requires further research and more granular transparency data.

All study data as well as the Python and R code used for analysis have been
released under a permissive license. Together with an extended technical report,
they have been archived in an academic research repository~\anon[{[\emph{elided
for review}]}]{\cite{Grimm2024}}.

\section{Methods}
\label{sec:methods}

To establish necessary context, this study starts out with a critical literature
review. It is based on a mix of broad and specifically targeted searches of both
internet and academic repositories that iteratively refine prior results. The
goal is to synthesize a realistic conceptual model of CyberTipline reporting
that captures the relationship between illegal, detected, reported, and uniquely
actionable incidents and offers granular detail even where existing transparency
practices do not. Since more than 99\% of all CyberTipline reports concern
\V{CSAM}, the literature review paid particular attention to breaking down this
monolithic category into subcategories reflecting kinds of \V{CSAM} observed in
practice. The iterative expansion of sources considered as part of the review
ensures that those subcategories are representative, though it cannot guarantee
that they are exhaustive, i.e., complete. Where data is available, we
incorporate it into the review. However, the review also surfaced several
aspects that would benefit from quantification.

\subsection{Data}
\label{sec:methods:data}

The subsequent parts of the study build on a new dataset comprising quantitative
\V{CSE} transparency disclosures for about thirty organizations, covering large
technology firms, their subsidiaries and social media brands, as well as a
couple of not-for-profits~\anon[{[\emph{elided for review}]}]{\cite{Grimm2024}}.
The latter include the National Center for Missing and Exploited Children or
\V{NCMEC}, which serves as clearinghouse for reports about \V{CSE}, which are
mandated by \V{US} law and received through its CyberTipline. Since social media
thrive on user-generated content, the primary criterion for inclusion in the
dataset---in addition to participating in CyberTipline reporting---is being a
popular social media site based on Buffer's yearly rankings from~2022
through~2024\TechReport{~\cite{Lua2022, Lua2023, Oladipo2024}}. To ensure broad
coverage, further criteria are being labelled a very large online platform by
the European Commission\TechReport{~\cite{EuropeanCommission2023b}} and having
filed 100,000 or more CyberTipline reports during any year, which are roughly
0.3\% of the total report count for 2023.

\begin{table}
\centering\libertineLF

\caption{An overview of surveyed organizations and their subsidiaries or brands.
Criteria for inclusion are social media (Soc), very large platforms (VL), and
platforms that filed $\geq100$k reports (.1M)}

\label{tab:survey}
\begin{tabular}{lcccl@{}rc}
\toprule
\textbf{Organization} & \textbf{Soc} & \textbf{VL} & \textbf{.1M} &
\multicolumn{2}{l}{\textbf{Disclosures (Months)}} &
\multicolumn{1}{l}{\textbf{Audit}} \\
\midrule
Amazon                &\yes&\yes&    & 2020..           & 12 & Reports \\
\hskip1.2em Twitch    &\yes&    &    & H1~2020..        &  6 &         \\
Apple                 &    &\yes&    &                  &    &   ---   \\
Automattic            &\yes&    &    &                  &    &   ---   \\
\hskip1.2em Tumblr    &\yes&    &    &                  &    &         \\
Aylo née MindGeek     &    &\yes&\yes&                  &    & Reports \\
\hskip1.2em Pornhub   &    &\yes&\yes& 2020..           &  6 &         \\
Discord               &\yes&    &\yes& H2~2020..        &  3 &   ---   \\
Google                &\yes&\yes&\yes& H1~2020..        &  6 & Reports \\
\hskip1.2em YouTube   &\yes&    &\yes& H1~2020..        &  6 &         \\
Meta                  &\yes&\yes&\yes&                  &    &  Data   \\
\hskip1.2em Facebook  &\yes&    &\yes& Q3~2018..        &  3 &         \\
\hskip1.2em Instagram &\yes&    &\yes& Q2~2019..        &  3 &         \\
\hskip1.2em WhatsApp  &\yes&    &\yes&                  &    &         \\
Microsoft             &\yes&\yes&\yes& H1~2020..        &  6 & Reports \\
\hskip1.2em LinkedIn  &\yes&\yes&\yes& H1~2019..        &  6 &         \\
\hskip1.2em Skype     &\yes&    &    &                  &    &         \\
\hskip1.2em Teams     &\yes&    &    &                  &    &         \\
NCMEC                 &    &    &    & 2019..           & 12 & Reports \\
Omegle                &    &    &\yes&                  &    &   ---   \\
Pinterest             &\yes&\yes&    & H1~2020..        &  3 & Reports \\
Quora                 &\yes&    &    &                  &    &   ---   \\
Reddit                &\yes&    &\yes& 2019..           &  6 & Reports \\
Snap                  &\yes&\yes&\yes& H2~2019..        &  6 & Reports \\
TikTok                &\yes&\yes&\yes& Q1~2022..        &  3 &  Data   \\
Wikimedia             &    &\yes&    &                  &    &   ---   \\
X                     &\yes&\yes&\yes& H1~2024..        &  6 &   ---   \\
\hskip1.2em Twitter   &\yes&    &\yes& H2~2018..H1~2021 &    &         \\
\bottomrule
\end{tabular}
\end{table}

\TechReport{Where available, an organization's entry in the dataset includes the
quantitative disclosures, preserving their numerical precision, unit names (with
some well-defined exceptions for common metrics), and reporting periods. At the
same time, some organizational entries record only the absence of relevant
transparency data, e.g., Apple, or the relationship to other entries that do
contain data, e.g., Meta's entry lists Facebook, Instagram, and WhatsApp as
corporate brands. }Table~\ref{tab:survey} lists the 15~corporations and
2~not-for-profits surveyed by this study, the criteria for inclusion, the
(indented) subsidiaries/brands meeting the criteria, time periods covered by
their transparency disclosures, current disclosure frequencies in months, and
audits applied to their data. Amongst the 15~corporations, Omegle is probably
the most obscure; the video chat service shut down in November 2023 to settle a
lawsuit filed by a minor who was sexually exploited on that platform. As shown
in the table, CyberTipline report counts were audited for seven corporations
plus \V{NCMEC}, and historical data disclosures were audited for another two
corporations.

\TechReport{
The dataset further includes several supplemental tables and code for more
easily processing the data. Supplemental tables comprise domain-specific
information, such as the original machine-readable disclosures by Discord, Meta,
Microsoft, and TikTok, as well as generally useful information, such as
per-country population counts from the \V{UN} Population
Division~\cite{WorldPopulationProspects2024} and global vector outlines from
Natural Earth.\footnote{\url{https://www.naturalearthdata.com}} Since Meta
updates the data but not \V{URL} for its quarterly \V{CSV} disclosures, the data
for \V{Q2}~2021 through \V{Q1}~2022 (inclusive) was accessed through the
Internet
Archive.\footnote{\url{https://web.archive.org/web/20240000000000*/https://transparency.fb.com/sr/community-standards/}}
While reported statistics go as far back as \V{Q3}~2018, no \V{CSV} files from
before \V{Q2}~2021 are available. To help clean up, validate, and enrich the
data, the code includes a 4,400-line Python library. It also includes the Python
and R notebooks with the analysis for this study.
}

As illustrated by the \emph{Disclosures} column in Table~\ref{tab:survey}, most
organizations release some transparency data. At the same time, \V{NCMEC}'s
disclosures are the by far most comprehensive, thanks largely to two reports to
the Congressional Committees of Appropriations for the years
2022~\cite{NcmecOjjdp2022} and 2023~\cite{NcmecOjjdp2023}. Mandated by the
explanatory statement for the Consolidated Appropriations Act of
2022\TechReport{~\cite{DeLauro2022}}, the data included in these disclosures
goes beyond the per-provider and per-country breakdowns published before and
provides, for the first time, comprehensive statistics on reported activities,
the uniqueness of reports and attachments, as well as the relationship between
perpetrators and abused minors. While most of the data covers recent years only,
i.e., from 2020 forward, the disclosures do include yearly CyberTipline report
counts going back to the beginning in 1998. These statistics are the foundation
for the second part of the study, which considers the volume of CyberTipline
reports.

\subsection{Audits}
\label{sec:methods:audits}

To assess the quality of technology industry transparency efforts, this study
includes two audits of transparency data as its third part. The first and
primary audit compares CyberTipline report counts for electronic service
providers or \V{ESP}s as disclosed by their producers and their only consumer.
While almost all service providers that disclose counts do so semiannually or
quarterly and several of them distinguish between subsidiaries or brands,
\V{NCMEC}'s yearly disclosure frequency and corporation-only breakdown for 2019
and 2020 determine the granularity for all audits, i.e., yearly per-corporation
counts. Out of the 16 organizations other than \V{NCMEC} contained in the
dataset, seven include CyberTipline report counts in their transparency
disclosures and hence are subject to the first audit. Twitter's transparency
disclosures only covered the number of unique accounts closed. However, X's
disclosures for \V{H1}~2024, for the first time, do include CyberTipline report
counts, which raises the possibility of inclusion in a future iteration of this
audit.

The second audit compares computer-readable data disclosures, e.g., \V{CSV}
files, by the same service provider for subsequent reporting periods. In
particular, for each reporting period $p+1$, it compares all entries shared with
the previous reporting period $p$, which should be the entries for periods $1$
to $p$ or all entries in the file for reporting period $p$. The dataset contains
two corporations that release current and past transparency data in
machine-readable formats and hence are subject to the second audit. They also
happen to be distinct from the seven corporations covered by the first audit.

Since most surveyed organizations engage in surveillance
capitalism\TechReport{~\cite{Zuboff2019}}, i.e., critically depend on their
ability to process and analyze large volumes of data, we might expect to mostly
find pairwise identical quantities. Furthermore, in the rare case that
quantities differ, i.e., exhibit some kind of error, we'd expect that either
service provider or \V{NCMEC} would have noticed and provided a credible
explanation with the data. Alas, in the real world, instances of divergent
transparency data aren't hard to come by, but apparent awareness and
explanations thereof are. In fact, we encountered only one instance of a service
provider explaining a major discrepancy. At the same time, the audits identified
significant data quality issues for three service providers. Furthermore, claims
made by two of the providers are sufficiently discordant with the data to raise
concerns about the veracity of their transparency reporting.

With numerical equality too blunt an audit instrument, the challenge becomes to
tease apart different kinds of errors. Hence we need to define what kind of
errors we can reasonably expect for each audit. For CyberTipline report counts,
we are comparing the results of two independent organizations tabulating the
same unidirectional message flows. It seems reasonable to allow for small errors
in the classification and tabulation of these messages due to
(non-deterministic) software bugs and hardware outages. Hence, differences
between comparable report counts should be small relative to their overall
magnitude, say less than 1\%, and not exhibit other discernable patterns, i.e.,
they reflect random errors only.

For data files, we are comparing the same metrics published by the same
organization some number of months apart. In this case, values should only
change as a result of deliberate modification, notably, to correct incomplete or
buggy disclosures. Hence, differences should exhibit temporal or logical
locality and hence qualify as systemic errors only. When considering both
audits, we note an altogether remarkable difference: While small random errors
are acceptable for the first audit, they are unacceptable for the second audit.
Likewise, while small systemic errors are acceptable for the second audit, they
are unacceptable for the first audit.

\subsection{Statistics}
\label{sec:methods:stats}

We did consider the intra-class correlation coefficient for count data for
comparing the tallies of CyberTipline reports~\cite{LCarrasco2022}. However, due
to the small number of per-year, per-provider samples, with $2\leq{}n\leq{}5$
and $\overline{n}=3.86$, we could only compute a goodness-of-fit for the entire
dataset, across all providers, with $N=27$. Those calculations demonstrate that
neither the Poisson distribution nor a negative binomial distribution with
variance growing linearly with mean were suitable models. A negative binomial
distribution with variance growing quadratically with mean fared better but
nonetheless couldn't account for all of the samples. Since the results offer
little insight beyond what can already be gleaned by inspection, we decided to
drop this marginal statistic.

Instead, we are basing our comparison on Bland-Altman or mean-difference
plots~\cite{BlandAltman1995} for each of the seven providers that publish
CyberTipline report counts as well as for all of them together. Consistent with
the above stated assumption that the error for CyberTipline report counts grows
with magnitude, we plot percentage differences, not absolute values. We use the
mean as denominator, since it provides an unbiased best guess for the true value
in the absence of other information. Since the normality assumption does not
hold for count data, we omit the customary 95\% limits of agreement from the
plots and instead trace the mean difference. We then rely on visual inspection
to analyze patterns and outliers, while also employing the sign test and
Fisher's exact test to confirm their statistical
significance\TechReport{~\cite{Kanji2006}}.

\subsection{Global Impact and Limitations}
\label{sec:methods:limits}

Even though this study is based on a mechanism mandated by \V{US} law, the reach
of American internet services all but ensures that the resulting data is of
global significance. Notably, Meta generated more than 84\% of all CyberTipline
reports every year since 2019 and more than 91\% from 2019 to 2021 (inclusive).
Furthermore, based on Meta's ``monthly active
people\Journal{,}''\TechReport{~\cite{FacebookInc2021, MetaPlatformsInc2022,
MetaPlatformsInc2024},} i.e., users that logged into Facebook, Instagram,
Messenger, or WhatsApp at least once during a month, and \V{UN} population
estimates\TechReport{~\cite{WorldPopulationProspects2024}}, Meta's user share
amongst the global population, across \emph{all} ages, grew from 37\% in 2019 to
49\% in 2023!

That reach, however, is not uniform, with China being the most significant
holdout. In fact, none of the social media platforms operated by the
corporations included in this study are accessible from within China.
\Journal{While typically not as comprehensive, several other countries similarly
restrict at least some American social media. }\TechReport{Though Microsoft's
LinkedIn did have a local equivalent, and TikTok, in addition to being under
Chinese ownership, does have a local equivalent. Furthermore, while Apple's
iCloud and iMessage are available within China, they use local servers that are
not operated by Apple. Amongst several others, Belarus, India, Iran, Pakistan,
and Russia similarly restrict at least some American social
media~\cite{FunkBrodyea2024}.

}Despite the uneven global reach, CyberTipline reporting has significant impact
on countries where American service providers have a substantial presence. For
example, in its transparency disclosures, Germany's federal police repeatedly
acknowledges \V{NCMEC}'s CyberTipline as the by far most significant source for
reports about \V{CSE}\TechReport{, though it does not quantify their
share~\cite{Bundeskriminalamt2024}. However, this study only considers
\V{NCMEC}'s per country statistics}. Analyzing the disclosures of international
police forces is an interesting and complimentary topic for future work.

This study focuses on centralized corporate social media because they currently
are the dominant means for accessing user-generated content. That is highly
effective, with the surveyed 16 out of almost 250 organizations accounting for
more than 97\% of all CyberTipline reports every year. Since these social media
are typically accessed through mobile devices, this study also includes the
currently dominant means for accessing \V{CSAM}~\cite{SteelNewmanea2020}.
\Journal{However, because it focuses on CyberTipline reports, this study has
less coverage for decentralized technologies, including the dark web,
peer-to-peer networks, and federated social media. That seems to be the case
even for federated social media that perform active content moderation. For
example, Mastodon gGmbH operates the largest Mastodon instance with 240,000
active users but filed a mere 16 CyberTipline reports in 2023. By comparison,
the bottom three organizations surveyed by this study are the Wikimedia
Foundation with 34 reports, Apple with 267, and Aylo with 2,596.}

\TechReport{To put it differently, this study only marginally covers formerly
dominant but still actively used technologies, such as peer-to-peer networks and
the dark web~\cite{SteelNewmanea2020}, as well as possible future dominant
technologies, such as federated social media. To begin with, most dark websites
and some peer-to-peer networks aren't particularly concerned with legal
compliance and thus highly unlikely to file CyberTipline reports. Meanwhile,
many federated social media instances may take their content moderation and
legal compliance more seriously, but they nonetheless won't (be able to)
allocate resources to trust and safety that are even remotely comparable to
centralized corporate social media. As a case in point, Mastodon gGmbH operates
the largest Mastodon instance with 240,000 active
users\footnote{https://mastodon.fediverse.observer/list} and also performs
content moderation. But even though \V{CSAM} is being traded across the
so-called fediverse~\cite{ThielDiResta2023}, the firm filed a mere 16
CyberTipline reports in 2023. By comparison, the bottom three organizations
surveyed by this study are the Wikimedia Foundation with 34 reports, Apple with
267 reports, and Aylo with 2,596 reports.}

That is not to suggest that the relationship between corporate social media and
the volume of CyberTipline reports can be reduced to anything as quaint as the
allocation of capital. On the contrary, the findings, particularly
Section~\ref{sec:findings:reports}, raise the distinct possibility that it took
the combination of surveillance-capitalist social media with their amplification
of base human emotions and potential for near-global virality, indiscriminate
and overreaching \V{US} laws, and cell-phone wielding teenagers to create
something akin to a perfect storm that keeps producing an ever worsening deluge
of \V{CSAM}. Remarkably, that same data also suggests that \V{CSE} is not nearly
as terrible a problem as \V{NCMEC}'s yearly statistics suggest.

\section{Background}
\label{sec:background}

\subsection{US Law}

Chapter 110 of 18 \V{US} Code\TechReport{~\cite{Chapter110Code18US}}, the United
States' criminal code, concerns the ``sexual exploitation and other abuse of
children.'' It prohibits grooming children (\S2251), selling or buying children
(\V{\S2251A}), the production, distribution, and possession of imagery depicting
minors engaging in ``sexually explicit conduct'' (\S2252) and any other
materials qualifying as ``child pornography'' (\V{\S2252A}) even in other
countries (\S2260). Further prohibitions apply to tricking minors into accessing
harmful materials with misleading domain names (\V{\S2252B}) or page contents
(\V{\S2252C}). As long as providers of electronic services immediately report
such activities and materials to \V{NCMEC}'s CyberTipline (\V{\S2258A}), it
explicitly limits their civil and criminal liability (\V{\S2258B}).

Penalties include criminal (\S2253) and civil forfeiture (\S2254) as well as
mandatory restitution (\S2259) in addition to fines and long prison sentences,
e.g., 5--20 years for the first violation of \V{\S2252A} and 15--40 years for
subsequent violations. The only violations of \S2252 and \V{\S2252A} without a
minimum penalty are possession of \V{CSAM}. According to the guidelines of the
\V{US} Sentencing Commission\TechReport{~\cite{ReevesMateea2024}}, a first-time
offender possessing less than 10 images, none of them involving prepubescent
children, sadomasochism, or violence, faces between 27--33 months in prison in
addition to a \$10k--\$100k fine. However, thanks to a 1996 amendment, if the
offense involves a computer, the penalty increases to 33--41 months in addition
to a fine of \$15k--\$150k. Still assuming less than 10 images, no prepubescent
children, \V{S/M}, or violence, if the offense involves offering/sending such
imagery to others, the penalty increases to 51--63 months. If it also involves
``pecuniary gain,'' ``valuable consideration,'' or a minor as recipient, the
penalty further increases to 87--108 months.

In addition to enshrining the reporting requirement for service providers into
law, Chapter~110 also prescribes \V{NCMEC}'s role as clearinghouse. Notably, it
compells the center to triage and forward CyberTipline reports to law
enforcement (\V{\S2258A}), authorizes it to collect image and video hashes and
share that information with service providers (\V{\S2258C}), and exempts the
organization from any civil and criminal liability that might arises from
running the CyberTipline (\V{\S2258D}). \Journal{Despite having been created
through a Congressional Act and receiving most of its funding from the \V{US}
federal government, \V{NCMEC} nonetheless is a private not-for-profit
corporation with an independent board and offers other, related services, such
as case management or family advocacy~\cite{FernandesAlcantaraHanson2021}.}

\TechReport{Despite having been created through a Congressional Act and
receiving most of its funding from the \V{US} federal government, \V{NCMEC}
nonetheless is a private not-for-profit corporation with an independent board.
As such, it offers other, related services, such as case management for missing
children, sex offender tracking, and family
advocacy~\cite{FernandesAlcantaraHanson2021}. Still, fulfilling its role as
clearinghouse necessitates close collaboration with law enforcement. Two federal
appeals courts have found that \V{NCMEC} qualifies as a government agent. Hence,
its inspection of CyberTipline report attachments during triage may amount to
illegal warantless searches, which may make evidence gathered during such
inspections inadmissible in court~\cite{GrossmanPfefferkornea2024}.}

\subsection{CyberTipline Reports}

Each CyberTipline report describes an individual incident. Accordingly, the
schema starts with a summary that identifies the type as well as date and time
of the incident\TechReport{~\cite{NCMEC2024}}. It is followed by sections that
capture granular information about involved internet services, interactions with
law enforcement, the reporter of the incident, the reported person, the intended
recipient, and the child victim. File attachments include violative materials
such as pictures and videos that qualify as \V{CSAM} as well as other
supplementary content.

A few fields are clearly designed to guide triage. The report summary includes
an \verb|escalateToHighPriority| flag and the \verb|sextortion|,
\verb|csamSolicitation|, \verb|minorToMinorInteraction|, and \verb|spam|
annotations. The metadata for attachments includes tags such as
\verb|potentialMeme|, \verb|viral|, \verb|possibleSelfProduction|,
\verb|liveStreaming| and \verb|generativeAi|. \TechReport{It also includes the
Tech Coalition classification\TechReport{~\cite{TechCoalition2022a}}. That
industry organization's labelling scheme distinguishes between
(A)~``pre-pubescent'' and (B)~``post-pubescent'' as well as between (1)~``sex
act'' and (2)~``lascivious exhibition'' resulting in the four labels \V{A1},
\V{A2}, \V{B1}, and \V{B2}.} A few actual CyberTipline reports (without
attachments) have become public as part of court
proceedings~\cite{CyberTipline5074778, CyberTipline5778397}.

\section{Findings}
\label{sec:findings}

We present this study's findings in three parts.
Section~\ref{sec:findings:law-and-tech} discusses the interaction between law
and technology, introducing a model for CyberTipline reporting that goes beyond
the simplistic narrative of sexual predation. When then shift focus to
quantitative analysis. Section~\ref{sec:findings:reports} addresses the growth
of CyberTipline reports over the last quarter century, which is best modelled as
two distinct epochs, the one before 2014 with very limited growth and the one
since 2014 with linear but aggressive growth. Section~\ref{sec:findings:audits}
presents the results of the two comparative audits. It demonstrates that seven
out of sixten firms are acting in good faith and releasing reasonably accurate
transparency data. One firm hasn't released enough data yet. The remaining eight
firms either are not releasing any data or have been making subpar data
disclosures. Amongst the second eight, Discord, Meta, and TikTok stand out as
particularly problematic.

\subsection{Law and Technology}
\label{sec:findings:law-and-tech}

\begin{finding}
The prohibitions of \V{\S\S2252--2252A} against sexual imagery involving minors
threaten severe punishment for activities that have little to do with sexual
predation. They increasingly theaten the very group they are intended to
protect, the minors themselves.
\end{finding}

\noindent{}Findings included with bills that updated chapter~110 tell a familiar
story. They emphasize a ``compelling State [sic] interest in protecting children
from those who sexually exploit them, and this interest extends to stamping out
the vice of child pornography at all levels in the distribution
chain\Journal{.}''\TechReport{~\cite{UnitedStatesCongress2008}.} They further
claim that ``technological ease, lack of expense, and anonymity \ldots [of] the
Internet has resulted in an explosion in the multijurisdictional distribution of
child pornography\Journal{.}''\TechReport{~\cite{UnitedStatesCongress2003}.}
They then position the criminal law as an effective means ``to dry up the market
for this material by imposing severe criminal
penalties\Journal{.}''\TechReport{~\cite{UnitedStatesCongress2008}.}

\V{CSAM} is indeed the by far most frequently reported incident type, accounting
for more than 99.2\% of all CyberTipline reports filed over the last four years.
However, there is scant evidence for an ``explosion'' of \V{CSAM}. As we will
demonstrate in Section~\ref{sec:findings:reports}, the primary driving force
behind a decade of rapid and \emph{linear} growth in CyberTipline reports is an
equivalent growth in the number of social media users.

At the same time, there is clear evidence that the majority of \V{CSAM}
incidents have little to do with sexual exploitation and tend towards the more
benign. In 2021, Meta developed a new taxonomy of intent for sharing \V{CSAM},
distinguishing between ``malicious'' and ``nonmalicious'' users. While the first
group intend to harm children, the second group ``are people whose behavior is
problematic and potentially harmful, but who we believe based on contextual
clues and other behaviours likely did not intend to cause harm to
children''~\cite{BuckleyAndrusea2021}. When applying that taxonomy to a sample
of 150~accounts that were reported to \V{NCMEC}, Meta found that malicious users
are clearly distinguishable from nonmalicious ones and that ``more than 75\% of
these did not exhibit malicious intent,'' including expressions of abhorrence,
inappropriate memes, and adolescents engaging in ``consensual, developmentally
appropriate sexual behavior'' by consensually sharing sexual imagery amongst
each other. The firm then leveraged these findings to roll out targeted
interventions~\cite{Davis2021}. A study based on semi-structured interviews with
trust and safety personnel, law enforcement officers, and \V{NCMEC} employees
has since found that particularly memes are leading to ``considerable
inefficiencies''~\cite{GrossmanPfefferkornea2024}.

While the prevalence of memes is not currently quantifiable, adolescents
consensually sharing sexual imagery, or ``sexting,'' is quantifiable and
ubiquitous. Recent metastudies have put the fraction of adolescents sending or
receiving sexts at, in publication order from 2018 to 2020 and 2022, 14.8\% and
27.4\%~\cite{MadiganLyea2018}, 14\% and 31\%~\cite{MollaEsparzaLosillaea2020},
as well as 19.3\% and 34.8\%~\cite{MoriParkea2022}. Given these rates, it is no
surprise that the lead author of the first study characterized sexting as a
``normative component of teen sexual behavior and
development''\TechReport{~\cite{Klass2018}}. Technically, teenagers who do sext
are violating federal and state laws against child pornography. While an
academic survey did attest significant restraint in prosecuting
adolescents~\cite{WolakFinkelhorea2012}, examples for (threatened) prosecutions
are not hard to come by\Journal{~\cite{Bekiempis2019, Feldman2020,
Jouvenal2023}}\TechReport{~\cite{Bekiempis2019, CrockettJr2016, Dunlap2016,
Feldman2020, Jouvenal2023, Miller2015, Moore2014, Tobias2009}}. Some states have
reduced the penalty for adolescent sexting to
misdemeanors\TechReport{~\cite{HindujaPatchin2022}}, but ``once they have the
option of lesser penalties, prosecutors are more likely to press
charges''~\cite{Hasinoff2016}.

While adolescent sexting can be a perfectly healthy exploration of sexuality, it
is not without significant risks. Notably, the slope from consensual to
nonconsensual sharing seems slippery, with the above metastudies reporting that
the prevalence for having forwarded a sext without consent is 12\%, 7\%, or
14.5\%, respectively. Additionally, having sexted correlates with being
subjugated to ``any image-based sexual abuse,'' increasing the prevalence from
2.8\% to 37.2\% amongst surveyed adolescents, a $13.2\times$
difference~\cite{FinkelhorSuttonea2024}. Minors engage in these behaviors even
though the vast majority ($>81\%$) know that doing so is illegal\TechReport{;
worries about legal consequences are especially acute for boys and Black
minors}~\cite{Thorn2020}.

\begin{figure}
\centering\libertineLF
\includegraphics[scale=0.4]{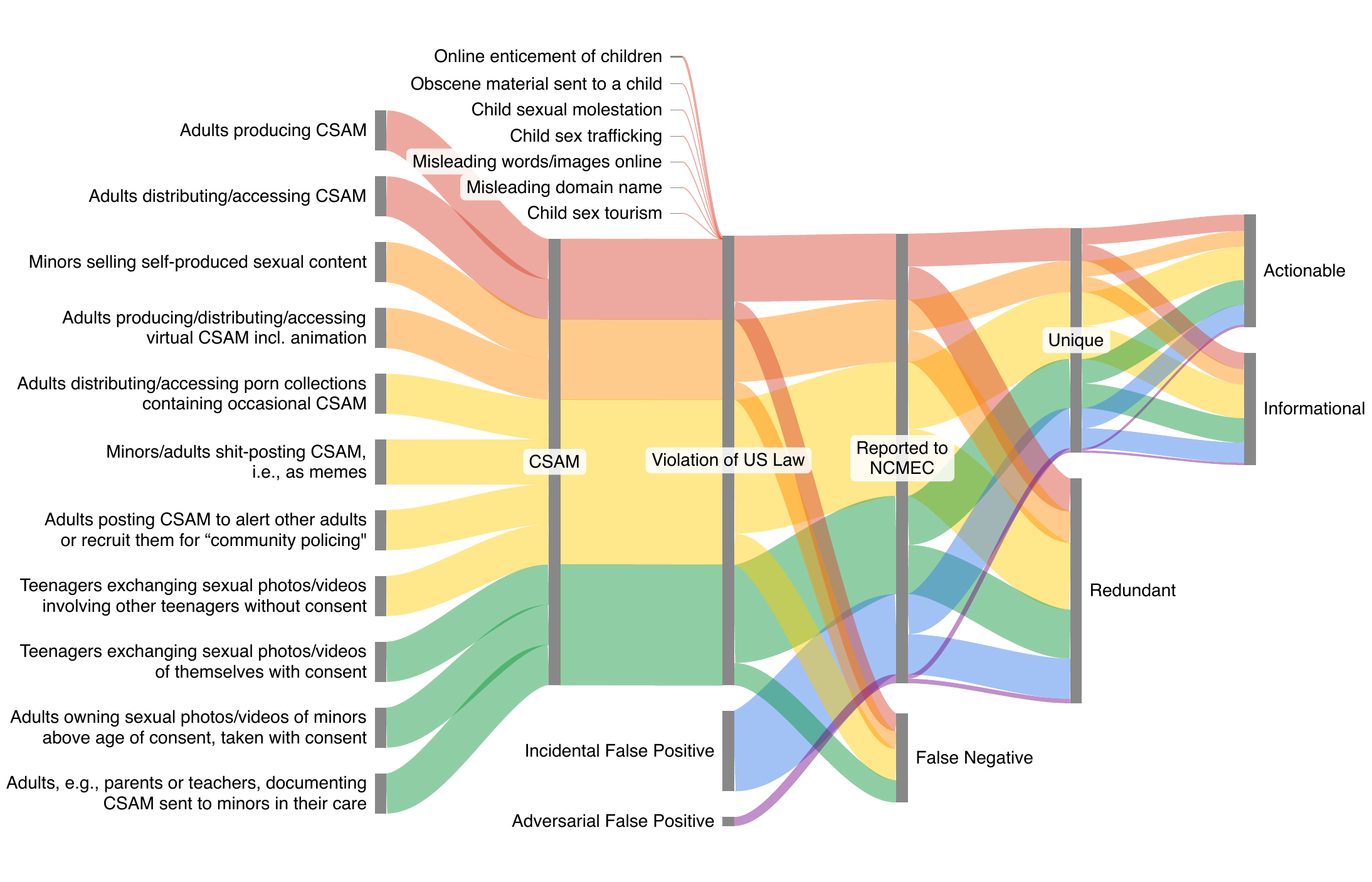}

\caption{The relationship between illegal, detected, reported, and uniquely
actionable incidents involving the sexual exploitation of minors. The
color-coding of \V{CSAM} subcategories from red to green represents a coarse
classification from directly harmful to mostly harmless.}

\label{fig:flow}
\end{figure}

Figure~\ref{fig:flow} illustrates the relationship between illegal, detected,
reported, and uniquely actionable incidents in CyberTipline reporting. The
\V{CSAM} subcategories on the left incorporate the types of incidents we just
discussed and several others encountered during the literature review. We
believe that the list is representative but make no claim of it being
exhaustive. \V{CSAM} subcategories are color-coded from red to green according
to a coarse classification of harm ranging from directly harmful down to mostly
harmless to minors. Flows for subcategories are sized arbitrarily because no
reliable statistics are available. Meanwhile, category-level flows into
``Violation of \V{US} Law'' \emph{are} sized according to observed CyberTipline
report ratios; almost all of them are \V{CSAM}.

\TechReport{
\begin{finding}
Most of the technologies used for detecting \V{CSAM} are vulnerable to
adversarial attack. \V{NCMEC} is still using \V{MD5}---16~years after \V{CERT}
advised to stop its use ``in any capacity.''
\end{finding}

\noindent{}Surveyed organizations rely on perceptual as well as cryptographic
hashes for detecting known \V{CSAM} and machine learning classifiers for
detecting previously unknown \V{CSAM}~\cite{LeeErmakovaea2020}. The critical
difference between perceptual and cryptographic hashes is that the latter
produce completely different values for inputs that differ by one bit only,
whereas the former produce similar values for similar inputs. That makes
perceptual hashes somewhat robust to changes in, say, image resolution or color
grading~\cite{Farid2021}. \V{NCMEC} uses both kinds, including the \V{MD5} and
\V{SHA-1} cryptographic hashes and Microsoft's PhotoDNA perceptual
hash~\cite{Farid2018}. With exception of Apple, Quora, and possibly Omegle, all
other organizations proactively scan user-uploaded content with some perceptual
hashing system, typically PhotoDNA. \V{NCMEC} and the Tech Coalition, an
industry group, maintain necessary hash databases, which are accessible to
industry.

In contrast to perceptual hashes, machine learning systems such as Google's
Content Safety \V{API} can detect previously unknown \V{CSAM}. Since possession
of \V{CSAM} is illegal, developing such machine learning models is more
complicated than for other image classifiers. Instead of directly training a
model on \V{CSAM}, it is necessary to combine independently and legally trained
classifiers for related traits starting with sexual activity and
age~\cite{LaranjeiraDaSilvaMacedoea2022}. Microsoft and Google make their
systems available for free to qualifying organizations.

In its 2023 transparency report to the Congressional Appropriations Committees,
\V{NCMEC} asserts that ``images that share the same \V{MD5} hash are
identical---as are images that share the same \V{SHA1}
hash''~\cite{NcmecOjjdp2022}. That is false. A successful collision attack
against \V{MD5}'s use in certificates has been demonstrated, and similar attacks
are now practical with consumer hardware~\cite{StevensLenstraea2012}. While the
computational cost of attacking \V{SHA-1} is substantially higher, collisions
for \V{SHA-1} and an attack on its use in \V{PGP} have been reported as
well~\cite{LeurentPeyrin2020a, StevensBurszteinea2017a}. Hence, the \V{US}
government's Computer Emergency Readiness Team (\V{CERT}) wrote in the closing
days of 2008 (!) that ``users should avoid using the MD5 algorithm \emph{in any
capacity}'' (emphasis ours)~\cite{ComputerEmergencyReadinessTeam2008}, and the
National Institute of Standards and Technology (\V{NIST}) started transitioning
away from \V{SHA-1} for \emph{all} applications in
2022~\cite{ComputerSecurityResourceCenter2022}.

Perceptual hashes don't fare any better. For many years, PhotoDNA benefitted
from security by obscurity. Since the algorithm was never published in
sufficient technical detail, it also wasn't subjugated to independent scrutiny.
More recently, however, PhotoDNA has been reverse-engineered and put through its
paces~\cite{Krawetz2021a}. As it turns out, PhotoDNA as well as Meta's
PDQ~\cite{Facebook2019} are vulnerable to efficient second preimage attacks,
which construct images that match other images' hashes, and also detection
avoidance attacks, which cloak an image so that it does not match its hash
anymore~\cite{ProkosJoisea2021}. On top of that and despite Microsoft's claims
otherwise, PhotoDNA is sufficiently reversible to recreate very small thumbnail
images~\cite{Athalye2021}. That raises the possibility that PhotoDNA hashes are
nothing other than cleverly encoded \V{CSAM}, which, of course, would make their
possession illegal.

Machine learning classifiers for images aren't more reliable either. A large
number of attacks have been documented in the research literature, including
both black box and white box attacks as well as attacks that falsely frame
victims and that cloak illicit materials~\cite{MachadoSilvaea2020}.
}

\begin{finding}
Automated scanning for \V{CSAM} has already been weaponized on Discord,
resulting in an unknown number of account terminations.
\end{finding}

\TechReport{
\noindent{}Turning the above, mostly theoretical vulnerabilities into practical
attacks and developing countermeasures are important topics for future work. But
if we were to speculate about real-world attacks, one credible scenario starts
with the leak of a hash database for PhotoDNA. Thereafter, those hashes are used
to stage a broad denial-of-service attack on the CyberTipline reporting system
by uploading borderline adult pornography modified to match leaked PhotoDNA
hashes. Alternatively, hashes are used in a more targeted manner: Through social
engineering, victims are manipulated into uploading seemingly innocuous images
that were modified to match leaked PhotoDNA hashes. This latter attack is akin
to swatting, but targets victims' internet accounts. Since many people rely on
the same few accounts for a large number of services, a false flag for \V{CSAM}
can have devastating consequences~\cite{Hill2022}.
}

\Journal{\noindent{}As described in the accompanying technical report, the image
hashing and machine learning technologies used for detecting \V{CSAM} are all
vulnerable to adversarial attack. That is gravely concerning given the
potentially devastating consequences of a false flag for
\V{CSAM}~\cite{Hill2022}. Worse, such attacks have already happened, as
illustrated by 2023 attack against Discord users~\cite{NoTextToSpeech2023}.}
\TechReport{The latter attack also provides the basic outline for a 2023 attack
against Discord users, except that no image modification was
necessary~\cite{NoTextToSpeech2023, Woohoowi2023}.} It involved manipulating
users on Discord into posting a specific image. The image seems innocuous
enough, showing an adolescent with headphones sitting in front of his computer
and eating popcorn. Reliably, when users posted the image, they did not receive
any promised rewards but their accounts were automatically terminated without
notification or explanation. Apparently, the still image belongs to a longer
video clip that shows the naked brother storming into the room and simulating
intercourse on camera. Since the clip qualifies as \V{CSAM}, Discord's trust and
safety team added the corresponding hashes to its database. But the firm was too
aggressive and included the innocuous beginning without the naked brother,
thereby creating an effective means of attacking gullible Discord users.

\TechReport{ Discussions on Reddit and YouTube about this weaponization of
falsely labelled \V{CSAM} employ a remarkable argot. Instead of using the term
``child pornography,'' most participants substitute the initialism ``cheese
pizza.'' While the term may have originated with sexual predators, it became an
important part of the PizzaGate conspiracy theory, and by now, it seems firmly
entrenched with the terminally online~\cite{Applebaum2016}.}

\begin{finding}
Reversal rates for content takedowns by Meta and account terminations by
Microsoft and Pinterest, after being actioned for \V{CSE}, have been slowly
increasing over the last few years and reached 3.1\%, 1.1\%, and 12.8\% in 2023,
respectively.
\end{finding}

\noindent{}Meta, Microsoft, and Pinterest are the only surveyed organizations to
release statistics about reversals of content takedowns (Meta) and account
terminations (Microsoft and Pinterest) for \V{CSE}. With exception of one out of
four years for Meta, the reversal rate has been slowly increasing for all three
firms, reaching 3.1\%, 1.1\%, and 12.8\% in 2023, respectively. Though, with
Meta's reversal rate dropping to 1.4\% over the first two quarters of 2024, it
looks like this trend won't continue for the firm. In any case, given the
significant potentila harm from being falsely accused of posting \V{CSAM},
Meta's and Microsoft's reversal rates are concerning, but Pinterest's rate is
outright alarming.

Figure~\ref{fig:flow} incorporates inadvertent as well as adversarial false
positives in the middle stages. It also shows false negatives, though no data is
available for the rate thereof. Likewise, we have no data on actual prevalence
of \V{CSAM}. Even Meta, which has been reporting prevalence rates for other
categories of violative content, has not published any statistics.

\begin{finding}
Additional factors that contribute to high CyberTipline report counts are
pictures and video that are reported more than once and low-quality reports
without sufficient detail to be actionable (called \emph{informational} by
\V{NCMEC}).
\end{finding}

\noindent{}As one would expect for actual abuse materials, memes, and image
shared in disgust, the same pictures and videos or \emph{pieces} are reported
more than once. The multiplication factor for unique imagery (as determined by a
cryptographic hash) and similar imagery (as determined by a perceptual hash) has
been decreasing over the last four years. In 2020, identical pieces were
reported $2.39\times$ on average and similar pieces $4.62\times$ on average. By
2023, those factors had come down to $2.09\times$ for identical pieces and
$3.10\times$ for similar pieces. The underlying reasons for this rather
substantial decrease in the degree of sharing over only four years are unclear.
They might include a fracturing of the online social fabric that (say, due to
polarization) reduces the reach of content in general, improvements in detection
technology that reduce the reach of \V{CSAM} specifically, or increasing
prevalence of content that legally counts as \V{CSAM} but isn't widely
distributed, e.g., sexts. Alas, determining the root causes will require
platform access to evaluate, for example, metadata.

In its transparency reports to the Congressional Appropriations Committees,
\V{NCMEC} remarks that the quality of CyberTipline reports differs
significantly. Hence, only about half of CyberTypline reports are actionable, as
opposed to what \V{NCMEC} euphemistically calls ``informational.''
Figure~\ref{fig:flow} captures the distinction between unique and redundant
attachments as well as actionable and informational reports as the final two
stages on the right.

\subsection{The Deluge of CyberTipline Reports}
\label{sec:findings:reports}

\begin{figure}
\centering\libertineLF
\includegraphics[scale=0.6]{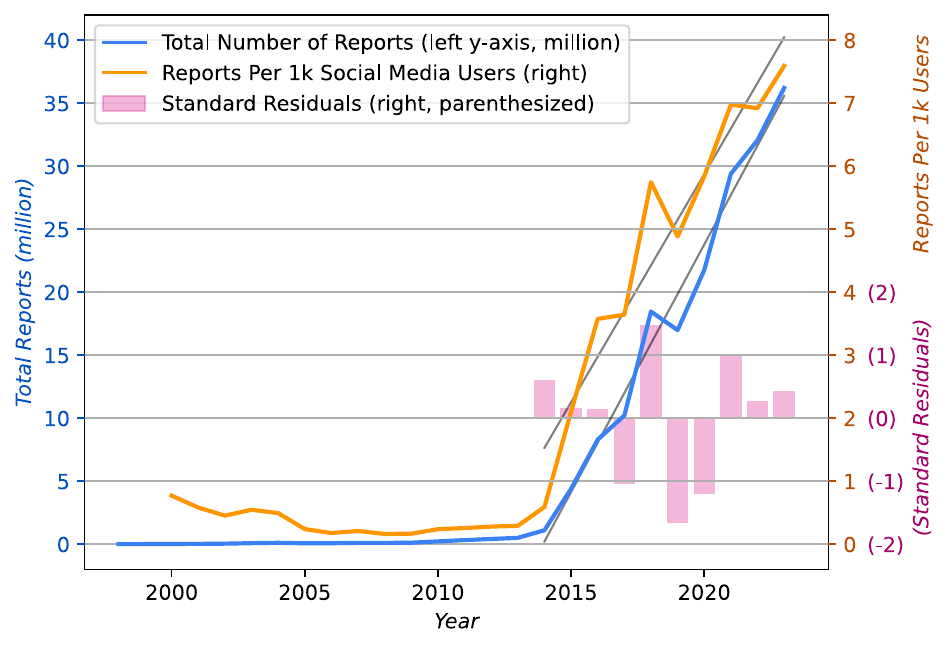}

\caption{Yearly CyberTipline reports (in blue, linear fit in gray, left y-axis),
reports per 1,000 social media user accounts (in orange, linear fit in gray,
right y-axis), and the standard residual (in magenta, right y-axis,
parenthesized).}

\label{fig:reports}
\end{figure}

Figure~\ref{fig:reports} shows three different views onto yearly CyberTipline
report counts, each with a different scale, as well as linear models for the
first two views. First, the lower, blue line tracks the absolute number of
reports in millions, with the left y-axis covering the range from~0 to~40
million reports. Second, the upper, orange line tracks the number of reports per
1,000 social media user accounts, with the non-parenthesized right y-axis
covering the range from~0 to~8 reports per 1k users. For both curves, gray lines
indicate the linear least squares fit for the decade since 2014 (inclusive).
Third, the magenta bars track the standard residuals for report counts during
that same decade, with the parenthesized right y-axis covering the range from~-2
to~2 standard deviations. Each of the three views illustrates one of the
following three findings.

\begin{finding}
Yearly changes in the number of CyberTipline reports are best modelled as linear
within two distinct epochs, with the first ending with 2013, the second starting
in 2014, and growth accelerating significantly.
\end{finding}

\noindent{}A 2019 study by authors affiliated with Google, \V{NCMEC}, and Thorn
characterized the growth of CyberTipline reports as
exponential~\cite{BurszteinBrightea2019}, and \V{NCMEC} has been doing the same
in its 2023 and 2024 transparency reports to
Congress\TechReport{~\cite{NcmecOjjdp2022, NcmecOjjdp2023}}. Alas, even in 2019,
an exponential regression wasn't a good fit for the data. With five more years
of data, it produces an even worse fit. Compared to the linear version, the
exponential fit has a $15\times$ larger mean squared error and the weighted
exponential fit has a $6.5\times$ larger error.

\begin{finding}
The likely primary driving force behind an absolute yearly increase of 3.93
million reports over the last decade is an equivalent growth in social media
user accounts. That does not, however, account for a relative increase from
roughly one to almost eight reports per 1,000 social media user accounts over
the same period.
\end{finding}

\noindent{}In retrospect, it is hardly surprising that social media user
accounts and CyberTipline reports are closely correlated. After all, social
media also file by far the most CyberTipline reports. This study, by design,
includes all major \V{US}-based social media. Amongst them, Meta alone filed
more than 92\% of all reports from 2019 to 2021 and 85\% thereafter. All social
media included in this study filed between 92\% and 96\% of reports from 2019 to
2021. Given that, it seems safe to assume that the relationship is causal: The
more users, the more users engaging in violative behavior or posting violative
content.

That contention is supported by the fact that related indicators do not exhibit
the same growth pattern as social media users. Notably, the number of internet
users has been growing either mostly linearly or vaguely exponentially over the
last 20 years, depending on source~\cite{ITU2023, RichieMathieuea2023}.
Interestingly, while the \V{ITU}'s global count of internet users shows mostly
linear growth over the last two decades, its count of \V{US} internet users from
2005 onwards, while far noisier than social media users or CyberTipline report
counts, shows the same tell-tale acceleration around 2014. That suggests that,
at least for the \V{US}, social media were the prime attraction for new internet
users over the last decade.

An analysis of \V{NCMEC}'s by-country
disaggregations\TechReport{~\cite{NcmecByCountry2019, NcmecByCountry2020,
NcmecByCountry2021, NcmecByCountry2022, NcmecByCountry2023}} offers additional
but qualified support. \TechReport{To begin with, the analysis surfaced a few,
minor data quality issues. Notably, we found entries for the same country under
two labels, ``French Guiana'' and ``Guiana, French,'' for the Netherlands
Antilles in addition to the three island states resulting from the Antilles'
breakup in 2010, for Bouvet Island even though this subantarctic Norwegian
territory is an uninhabited nature preserve, and for ``Europe'' in addition to
pretty much all European countries. With all but one count in the single digits,
we simply discounted these entries for the per-capita analysis.

}When ranking countries by reports per capita, we discovered an initially
puzzling phenomenon: From 2019 to 2023, not only was either Libya or the United
Arab Emirates (\V{UAE}) the country with the most reports per capita, but member
countries of the Arab League accounted for 13, 13, 13, 9, and 14 of the
top-twenty spots of countries with the most reports per capita. The substantial
differences between the two top-placed countries---with Libya spending much of
the 2010s in a state of civil war (ironically made worse by the \V{UAE}'s
interference) and the \V{UAE} the far more stable and denser populated country
with better infrastructure and larger \V{GDP}---only added to the puzzlement.

DataReportal's statistics on global social media penetration point towards a
simple explanation. Their ranking of countries with the most social media user
accounts per capita for 2023 shows seven Arab League countries including Libya
and the \V{UAE} amongst top-ten countries\TechReport{~\cite{Kemp2024}}. That
neatly explains the relatively high per-capita report count for that year.
However, with five and three Arab League countries amongst the top-ten for 2022
and 2021\TechReport{~\cite{Kemp2023, Kemp2022}}, the numbers do not fully align
with the reports per capita. Unfortunately, DataReportal publishes only top-ten
and bottom-ten countries. Hence we do not have enough information to determine
whether this is a data quality issue or reflects the impact of another,
unidentified variable.

Social media user counts may be the driving force behind the increase in
CyberTipline reports, but it is not the only one. As illustrated by the orange
curve in Figure~\ref{fig:reports}, normalized report counts have grown from less
than 1~report per 1,000 users to almost 8~reports per 1,000 users over the last
decade. That trend is certainly concerning, and identifying its cause is
critical future work.

\TechReport{
By comparison, incident rates for offline \V{CSE} in the \V{US} reportedly are
around 1--2 incidents per 1,000 \emph{children} (not
capita)~\cite{RezeyDiMeglio2024}. With children making up 22\% of the American
population~\cite{FederalInteragencyForumOnChildAndFamilyStatistics2023}, that
corresponds to 1--2 incidents per 4,550 capita or 0.22--0.44 incidents per 1,000
capita. When using the 29\% fraction of the global population as a scaling
factor instead~\cite{WorldPopulationProspects2024}, the offline rate would be
1--2 incidents per 3,450 capita or 0.29--0.58 incidents per 1,000 capita.

Since the two rates are normalized by different quantities, they are not
directly comparable. First, due to \V{US} privacy laws, most social media do not
allow children under 13 to open accounts. Even if this minimum age is not
strictly enforced and widely violated, children under 13 are going to be
underrepresented amongst social media users when compared to the general
population. Second, people may have accounts on more than one social media
platform as well as multiple accounts per platform. For instance, one survey
reports that 38\% of girls 13--17 and 37\% of \V{LGBT} kids have secondary or
``Finsta'' accounts~\cite{Thorn2020}. Assuming that the undercounting and
overcounting factors cancel each other out, the online prevalence of \V{CSE}
appears to be an order of magnitude larger than offline activities. Unless we
are inclined to believe that social media turn people into sexual predators at
significant rates, that difference must be due to other factors, which is
consistent with findings in the previous section.
}

\begin{finding}
The Covid-19 pandemic had no statistically relevant impact on the volume of
CyberTipline reports.
\end{finding}

\noindent{}As illustrated by the magenta bars in Figure~\ref{fig:reports}, the
standard residuals for CyberTipline reports are not particularly large, ranging
from~-1.7 to~1.5 standard deviations. In other words, the data is noisy when
compared to its model, but there are no pronounced outliers. Furthermore,
with~-1.2 and~+1.0, the standard residuals for the pandemic years 2020 and 2021
are clearly smaller, i.e., well within the noise. Finally, when considering
year-over-year growth, the increase of +0.5 standard deviations from 2019 to
2020 actually was on the smaller end. In other words, the growth in CyberTipline
report volume for the pandemic years is consistent with the growth for the rest
of the decade.

We could construct more complex scenarios under which the pandemic significantly
reduced organizations' report processing capability and hence a relative
increase in incidents was offset by a corresponding decrease in filed reports.
But that seems unlikely because many of the claims in support of a more
sustained impact do not hold up to scrutiny. Notably, besides a blog post from
April 2021 by \V{NCMEC} that features only year-over-year
comparisons~\cite{ODonnell2021} and a university's research report from May 2021
that captures the pressures faced by individual hotline staff and police
officers during the pandemic~\cite{SalterWong2021}, peer-reviewed
articles~\cite{Dabrowska2021, Haney2021b, SalterWoodlockea2023} consistently
cite reports by Europol from June 2020~\cite{Europol2020} and Interpol from
September 2020~\cite{Interpol2020}.

Both reports mostly feature anecdotal and indirect evidence. Both reports also
misrepresent data on peer-to-peer networks by the Child Rescue Coalition. They
claim that graphs of sinusoidal signals that happen to rise between February and
April 2020 somehow demonstrate a ``significant increase in the sharing of
\V{CSEAM}'' (child sexual exploitation and abuse material)---to use Interpol's
boldface phrase. In addition to sharing the same sourcing and description,
Interpol's single curve very much looks like the sum of Europol's four
country-specific curves. That suggests that both police agencies are presenting
the same data. However, according to y-axis labels, Interpol's counts are
roughly $5,700\times$ larger than Europol's aggregates.

At the same time, Europol's report does contains compelling evidence that the
pandemic did, in fact, result in significantly more incidents involving
\V{CSAM}, albeit \emph{at a different timescale}. When looking at CyberTipline
reports \V{NCMEC} forwarded to European police during the first half of 2020 at
\emph{monthly} granularity, there is a $5\times$ spike for March 2020 only.
While we concur that the pandemic is the most plausible cause, the effect is not
sustained enough to have any impact on yearly statistics. Alas, since we didn't
follow a formal protocol such as
\V{PRISMA}\TechReport{~\cite{MoherLiberatiea2009a}} in selecting publications,
we cannot exclude confirmation bias for the selection of the above examples.

\TechReport{
Before we shift focus to the two quantitative audits, we note the following
negative finding:

\begin{finding}
Whereas the total number of CyberTipline reports per year is a reasonably good
match for a linear model, that is not the case for most providers' report counts
nor for the number of attachments.
\end{finding}

\begin{figure}
\centering\libertineLF
\includegraphics[scale=0.6]{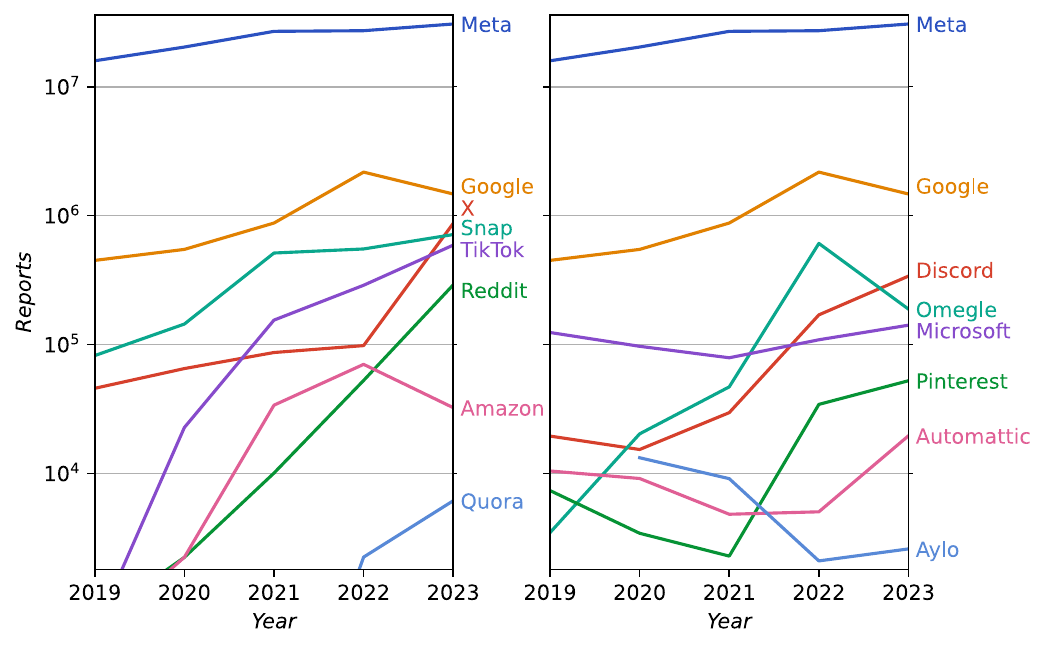}

\caption{Yearly CyberTipline report counts disaggregated by platform on a log
scale.}

\label{fig:platform}
\end{figure}

\noindent{}Figure~\ref{fig:platform} shows the log of yearly, per-provider
report counts across two equally dimensioned grids. Report counts for Meta and
Google are large and distinct enough to cleanly fit into both grids. By
contrast, report counts for Apple and Wikimedia are too small to fit into the
grids' 1,800--36,000,000 reports range. All other providers appear in one of the
two grids only, with placement based on a subjective judgement about
readability. As the figure makes clear, per-provider report counts show
significant variability and aren't even consistently growing. Out of four
providers with consistently growing report counts, Reddit has seen the most
aggressive, that is, at least exponential growth.

\begin{figure}
\centering\libertineLF
\includegraphics[scale=0.6]{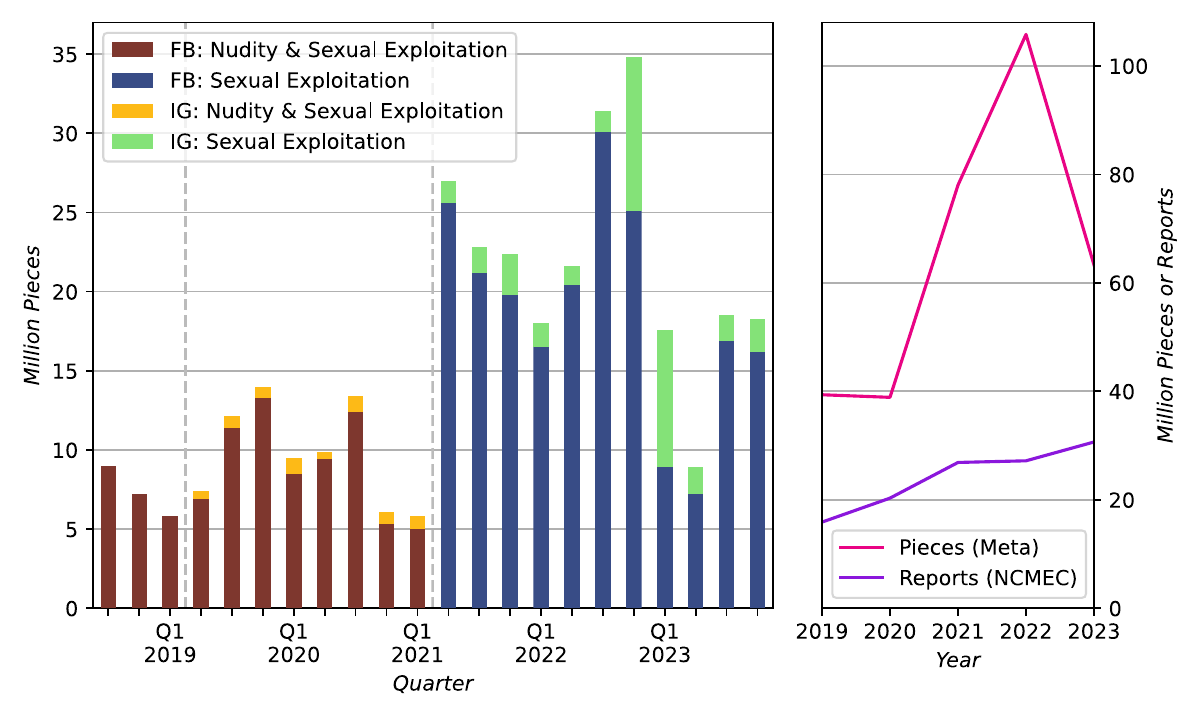}

\caption{Meta: The number of photos and videos, i.e., pieces, removed per
quarter on the left; number of pieces and reports per year on the right.}

\label{fig:meta}
\end{figure}

To characterize attachment counts, we turn to the most prominent filer of
CyberTipline reports, Meta. The left panel of Figure~\ref{fig:meta} shows the
firm's quarterly counts of images and videos taken down because they were
classified as ``child nudity \& sexual exploitation'' until \V{Q1}~2021
(inclusive) and as ``child endangerment: sexual exploitation'' thereafter. The
right panel shows those same piece counts aggregated by year and juxtaposed with
report counts, as disclosed by \V{NCMEC}.

At first glance, the change of metrics in \V{Q2}~2021 is curious because counts
jump upwards, even though the label suggests a smaller category. Meta's
transparency report for \V{Q2}~2021 provides no explanation. However, searching
for ``Meta Community standards enforcement report \V{Q2}~2021'' with an external
search engine did produce a blog post\TechReport{~\cite{Facebook2021a}} linking
a \V{PDF} document with the explanation~\cite{Facebook2021}. Apparently, the new
metric covers not only ``sexual exploitation,'' but also ``sexualization of
children'' and ``inappropriate interactions with children,'' which were not
tracked before. While all three are clearly related to minor safety, their
aggregation make little sense. But it does explain the jump in actioned images
and videos.

Even though Meta generated more than 91\% of all CyberTipline reports in
2019--2021 and more than 84\% in 2022--2023, the number of actioned pictures and
videos does not reflect the same consistent growth year-over-year. Particularly
striking are the variability between individual quarters and the substantial
yearly drop in actioned pictures and videos from 2022 to 2023. One possible
reason is that not all pictures and videos actioned in 2022 were severe enough
to warrant reporting to \V{NCMEC}. However, even if that's the case, the large
variability is surprising. While Meta announced its intent to share more
statistics about minor safety earlier this year, so far its follow-through has
been lacking---and since its transparency reports are not archived, that promise
appears to be gone, too.
}

\subsection{Transparency Audits}
\label{sec:findings:audits}

\subsubsection{Providers vs NCMEC}

\begin{figure}
\centering\libertineLF
\includegraphics[scale=0.7]{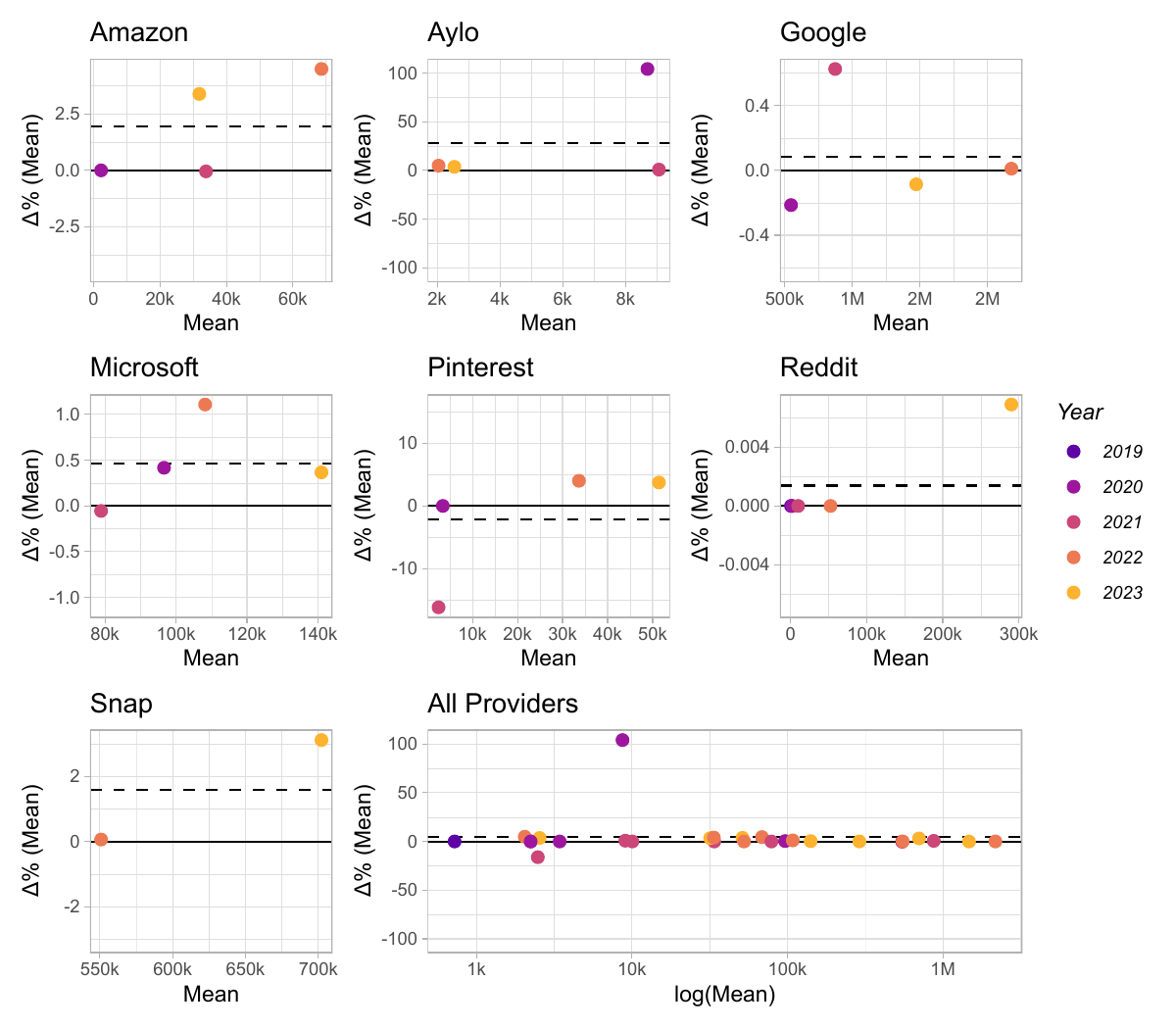}

\caption{Mean difference plots for seven service providers individually and all
of them together. Each plot has its own scale, and the x-axis is logarithmic for
the eighth plot.}

\label{fig:reports-audit}
\end{figure}

Figure~\ref{fig:reports-audit} shows the mean difference plots for the seven
service providers that released CyberTipline report counts, first individually
for each provider across seven panels and then for all providers in the eighth
panel. For each pair of report counts disclosed by a service provider and
\V{NCMEC}, the mean difference plots include a point with $x$ the mean of the
two counts, $y$ the difference of the two counts as a percentage of the mean,
and the color reflecting the year. If the service provider's count is smaller
than \V{NCMEC}'s, the percent difference is positive (and vice versa). The
dashed line is the mean of the percentage differences. Each of the eight plots
has its own dimensions for axes and the eighth plot's x-axis is logarithmic.
\TechReport{
Tables~\ref{tab:pieces-and-reports-1}--\ref{tab:pieces-and-reports-3} in
Appendix~\ref{sec:appendix:data} provide the raw data.}

\begin{finding}
Aylo's Pornhub is the only service provider to acknowledge and explain a mistake
in their handling of CyberTipline reports.
\end{finding}

\noindent{}When considering y-axis scales, the plot for one provider sticks out:
Aylo's 104.1\% difference for 2020. The firm only makes transparency disclosures
for its most visible brand, Pornhub. In its transparency report for the year,
which was published in April 2021, the brand not only acknowledges the
discrepancy but also explains it as the result of submitting the same reports
``multiple times in an abundance of caution''\TechReport{~\cite{Pornhub2021}}.
They add that \V{NCMEC} has been notified of the duplicate reports. Strikingly,
this is the only instance we could find of a service provider acknowledging a
problem with their disclosures or providing a credible explanation for the
problem. While we consider the acknowledgment and explanation exemplary, the
rest of the report makes for tedious reading. It emphasizes, again and again,
Pornhub's principled stand against any form of abuse.

\TechReport{Presumably, the firm is still reacting to events from December 2020,
when New York Times columnist Nick Kristof~\cite{Kristof2020} made common cause
with anti-porn crusaders~\cite{Hitt2020a} associated with a White supremacist
church~\cite{Halley2021} and used a few harrowing cases to falsely paint Pornhub
as the epitome of abusive sleaze~\cite{Brown2020, Grant2020}, a site ``infested
with rape videos.'' No matter that similar cases regularly occur on social media
as well. No matter that Pornhub had been working with \V{NCMEC} since 2019 and
was proactively scanning user-contributed content already. No matter that Apple
is not scanning proactively to this day. No matter that Kristof had published
similarly sensationalist and ultimately false reports twice
before~\cite{Dickson2014, Soderlund2011} and keeps casting himself in the role
of the superhero rescuing damsels from monsters~\cite{Heber2024}.}

\begin{finding}
For the seven service providers that disclose at least a year of CyberTipline
report counts, almost all of their counts are within 5\% of \V{NCMEC}'s
disclosures. They also exhibit a very small but statistically relevant bias
towards underreporting.
\end{finding}

\noindent{}Other than Aylo's outlier and a considerably smaller one at -16.1\%
for Pinterest, all percent differences have a magnitude smaller than 5\%. That
is larger than the 1\% threshold we set before data analysis (see
Section~\ref{sec:methods:audits}). But it still is reasonably small. While that
points towards a good faith effort at data collection and disclosure by those
seven service providers, the mean difference plots make another pattern clearly
discernible, namely that the majority of percentage differences is positive.
That holds for individual values, with 16 positive, 6 zero, and 5 negative
percentage differences. That also holds for the means of the percentage
differences, which are indicated by a dashed line in each panel and positive for
six out of seven providers as well as for all providers together. Even when
disregarding the two outliers, the remaining 25 percent differences still have a
positive mean of 1.2\%. The sign test further validates that this bias is
statistically significant with a p-value of 0.0133.

In theory, positive percentage differences are either the result of providers
undercounting or \V{NCMEC} overcounting. In practice, \V{NCMEC} has little to
gain by inflating report counts. After all, its role as clearinghouse for
CyberTipline reports is legally mandated, and it receives federal funding for
that function. By contrast, service providers have strong reputational and
marketing incentives to make their platforms appear ``safe,'' lest they upset
advertisers or (European) regulators~\cite{Roth2022}. It would appear that these
pressures are reflected in the data. However, at 1.2\% on average, the effect is
small enough so it doesn't distort the data in a meaningful way.

\subsubsection{Providers Over Time}
\label{sec:findings:audit:files}

For the previous audit, which compared counts tabulated by independent
organizations, a small amount of random error is not really surprising. For
this, the second audit, that is emphatically not the case. In fact, we only
considered the possibility after serendipitously using Meta's \V{CSV} file for a
subsequent quarter while validating a code refactoring. Except it didn't feel it
at the time, as we spent the better part of a day checking and rechecking our
code for bugs---until finally concluding that it had to be the input.

\begin{finding}
In addition to only including rounded quantities, Meta's transparency data is
marred by seven consecutive quarters with an average of 3.6\% divergent entries
over the previous quarter. Divergent entries vary widely in metrics and
quarters, with no apparent temporal or logical locality.
\end{finding}

\noindent{}For seven quarters, from \V{Q3}~2021 to \V{Q1}~2023 (inclusive),
Meta's \V{CSV} file diverges in an average of 3.6\% and range of 0.6\%--7.3\% of
all entries from the previous quarter's \V{CSV} file. Divergent entries vary
widely in policy area and actual metric (e.g., content actioned, appealed, or
restored). Notably, 12, 3, 7, 7, 11, 17, and 4 entries about \V{CSE} differ from
the previous quarter over that time range. Divergent entries also vary widely in
the affected period, going back up to 7--11 quarters every quarter but
\V{Q4}~2021 and \V{Q1}~2023. The latter two quarters also are the quarters with
the fewest divergent entries of 1.2\% and 0.6\%, respectively. Overall, there is
no discernible temporal or logical locality to these errors. Meta never
acknowledged these discrepancies. Though they did stop just as the \V{EU}'s
Digital Services Act was taking effect.

Worse, the actual extent of divergent entries across Meta's transparency data is
unknowable. Since Meta is the only service provider to round its transparency
data, additional disparities may very well have been hidden by the rounding
process. By the same argument, we cannot exclude that Meta's data continues to
suffer from quality issues. They may just be small enough not to survive
rounding. The presence of disparities also raises the question of why Meta
decided to round transparency statistics in the first place.

Meta fared better in two previous audits of its ``community standards
enforcement report,'' which the firm initiated itself and which did not find any
significant short-comings. First, a panel of academic experts performed a
qualitative review of Meta's policies, enforcement process, and transparency
reporting including metrics~\cite{BradfordGriselea2019}. Second, Ernst \& Young
performed a quantitative assessment of Meta's transparency data collection
according to usual accounting standards\TechReport{~\cite{Meta2022}}. According
to \V{E\&Y}'s written statement, the accountants reviewed Meta's management
assertion about the transparency data and found that ``internal controls were
suitably designed, implemented and operating with sufficient effectiveness.''
Now, \V{E\&Y} only reviewed \V{Q4}~2021, and that quarter had the second lowest
rate of divergent quantities, with $27/2,192=1.2\%$ of data points. Still, their
assessment clearly wasn't sufficiently thorough.

While Meta has certainly invested significant resources into trust and safety,
the firm also has a substantial history of faulty and misleading metrics and
favoring profits over minor safety. Notably, it misreported advertising metrics
to customers\Journal{~\cite{BruellPatel2020}}\TechReport{~\cite{BruellPatel2020,
Hutchinson2016, Hutchinson2016b, Hutchinson2017, VranicaMarshall2016}}, platform
usage data to a Harvard-based research consortium~\cite{Timberg2021}, and
adverts in its transparency
database\Journal{~\cite{Rosenberg2019}}\TechReport{~\cite{Rosenberg2019,
ScottMontellaro2021, SilvermanMac2020a}}. Meta also deplatformed the academics
who collected the data behind several of the news reports just
cited\Journal{~\cite{EdelsonMcCoy2021a}}\TechReport{~\cite{EdelsonMcCoy2021,
EdelsonMcCoy2021a}} and shut down its CrowdTangle transparency tool when it
became inconvenient~\cite{Kerr2024}.\anon[]{ While employed by Meta, the author
also discovered compelling evidence that the firm was artificially inflating
basic advertising metrics~\cite{Grimm2022c}.} As internal documents unsealed as
part of a lawsuit by the State of New Mexico against Meta
demonstrate~\cite{SungSilberling2024}, Meta running an advertising campaign (for
TikTok nonetheless) that sexualized teenage girls for the benefit of middle-aged
straight men was not an isolated incident~\cite{SilvermanMac2020}.

\begin{finding}
TikTok's transparency data on \V{CSE} is mostly unusable. Its \V{CSV} files are
marred by differences in label casing and floating point serialization. Claims
made in its \V{Q2}~2023 transparency report about improved disclosure
granularity are false when it comes to minor safety.
\end{finding}

\noindent{}TikTok discloses absolute counts only for the number of actioned
videos; all other metrics such as the breakdown into categories and
subcategories of violative content are expressed as decimal fractions. However,
before the firm's introduction of a new schema in \V{Q2}~2023, the firm did not
disclose all subcategory fractions, notably omitting the fraction of videos
actioned by automation, and therefore made it impossible to calculate the number
of videos actioned because they contain \V{CSAM}. The firm claimed that the new
schema ``added granularity.'' With an increase from 26 to 29 subcategories, that
is technically true. But in practice, TikTok also removed granularity, notably
reducing the number of subcategories for minor safety from five to four, which
are too coarse, again making it impossible to determine the number of videos
actioned because of \V{CSAM}. Meanwhile, the firm's \V{CSV} files are marred by
inconsistent capitalization for labels and inconsistent serialization of
floating point numbers, e.g., the same quantity may appear as
``0.9620000000000001'' and ``0.962.''

The lack of complete statistics when it comes to \V{CSE} suffices for rendering
TikTok's transparency disclosures useless for our purposes. However, even if it
was disclosing the necessary metrics, the subpar state of their \V{CSV} files
would unnecessarily complicate analysis.

\section{Discussion}
\label{sec:discussion}

Knowing that much of the growth in CyberTipline reports is driven by an
equivalent growth in social media users doesn't make \V{NCMEC}'s work any
easier. The organization is already close to breaking
point~\cite{GrossmanPfefferkornea2024}. But at the current rate of 3.93 million
more reports per year, it will need to process an additional 7.5~reports for
every minute in every year. At some point in the future, growth will necessarily
plateau. After all, there are only so many humans who can maintain only so many
social media accounts. Alas, assuming that children under 10 mostly stay off
social media and the rest of humanity maintain two social media accounts on
average (compared to six for \anon[one of ]{}this study's author\anon[s]{}),
that still implies $2.74\times$ the current CyberTipline report volume. Hence
reducing the volume of CyberTipline reports is imperative.

\subsection{Legal Overreach}
\label{sec:discussion:legal}

This study suggests two avenues for meaningfully reducing the CyberTipline
report volume. The first is legislative reform based on the realization that, in
the majority of cases, sexual imagery involving minors does not involve sexual
predation. That makes the blanket prohibition against such imagery excessively
punitive. It also impedes the self-determination of adolescents. While the
United States is one of only two countries in the world that hasn't ratified the
United Nations Convention on the Rights of the Child~\cite{Mason2005} and thus
does not recognize freedom from excessive interference as a right for children,
the blanket prohibition does interfere with other \V{US} laws.

For instance, in roughly half of \V{US} states, the age of consent is 16 or 17.
By definition, that allows for consensual sexual relationships between an
adolescent and an adult. However, if the adult takes (consensual) pictures of
their interactions and shares those pictures with their partner, the adult not
only violates the prohibition against \V{CSAM} but also meets the Sentencing
Guidelines' conditions of sending violative images to a minor, thus facing a
\emph{minimum} penalty of 87--108 months in prison---for privately sharing
pictures of a consensual and legal activity with the other participant. All it
takes for police to get involved is for one of the minor's parents, who happens
to disprove of the relationship, to look through their child's cell phone
pictures. That threat is particularly acute for \V{LGBT} youth, who encounter
substantial discrimination in the criminal injustice system~\cite{Silhan2011}.

Unfortunately, \V{NCMEC} and other advocacy groups are aiding this legal
overreach by referring to sexual imagery created by and involving minors as
``self-generated child sexual abuse materials'' or shorter ``self-generated
\V{CSAM}.'' It strongly suggests that these organizations are unable to conceive
of adolescents as sexual beings just when they are particularly focused on just
that aspect of their rapidly maturing bodies. That, in turn, gets in the way of
supporting adolescents in navigating what activities are and are not acceptable,
notably, when sexting. When all imagery is verboten, working out the practical
(and moral) differences becomes much harder. The fact that non-consensual
sharing has a substantial prevalence $>10\%$ suggests that distinguishing
between the two is, in fact, challenging for many adolescents.

\emph{None} of the above is an argument for legalizing the public posting or
arbitrary exchange of sexual imagery involving minors. But these scenarios serve
as compelling arguments for \emph{selectively} eliminating the attendant
criminal jeopardy. It may also be worth considering whether the exchange of
inappropriate memes or expressions of abhorrence are better treated as something
akin to traffic violations. Since the mid-1990s, the \V{US} spent about
\$2~billion on abstinence-only sex education, with nothing to show for that
investment besides ``increasing adolescent birthrates in conservative
states''~\cite{FoxHimmelsteinea2019}. It would be pure folly to keep trying this
same failed strategy for sexting.

\subsection{Wither the Deluge}
\label{sec:discussion:deluge}

The second second avenue for reducing the deluge of reports is undoing the
growth of normalized yearly rates from one to eight reports per 1,000 user
accounts. Compared to the other statistics in this paper with their millions to
thousands of millions, these are human-conceivable dimensions. While seemingly
larger than previously reported rates for offline \V{CSE}, they nonetheless are
inconsistent with the notion that the internet activated a previously latent
human potential for sexual predation. Still, the increase from one to eight
reports per 1,000 user accounts over the last decade is very clearly noticeable.

Making sense of this component requires additional research---and more granular
data including on the prevalence of sexts, memes, and predatory materials.
However, several factors point towards social media being responsible. There is
the finding that most of the growth is driven by social media user accounts.
There is the ``move fast and break things'' mindset of \V{US}-based corporate
social media and their relentless persuit of hypergrowth. Not surprisingly, the
resulting platforms make posting of user-generated content easier than any other
platform before and are designed for the viral distribution thereof. Facebook
contributing to the genocides in Myanmar and Ethiopia are extreme examples for
what can go wrong as a result\TechReport{~\cite{Gilbert2020, GlobalWitness2022,
HumanRightsCouncil2018, Mozur2018}}. Finally, there is commentary by observers
of internet culture that attests to 2014 as the year when social media displaced
the open web\Journal{~\cite{Cao2024}}\TechReport{~\cite{Cao2024, Staltz2017}}.

While suggestive, more work is clearly needed to make sense of this phenomenon
let alone undo it. In the meantime, more detailed reporting by the technology
industry can make a significant difference so that \V{NCMEC} can continue
serving as an effective global clearinghouse for reports about \V{CSE}.

\TechReport{
\subsection{Further Failures of Transparency}

In addition to the audits, the process of selecting and evaluating service
providers in preparation for this study surfaced concerns about the transparency
practices of a few more firms.

When Discord first disclosed statistics about its CyberTipline reporting in its
transparency report for July to December
2020\TechReport{~\cite{DiscordSafety2021}}, it did so in a section titled
``Reports to \V{NCMEC}'' in a bar chart titled ``\V{H2}~2020 \V{NCMEC}
Reports.'' Based on this labelling, we started tracking Discord's statistics as
CyberTipline reports. Subsequent transparency reports did not use the same
section headers, but still labelled the dominant statistic in their breakdown of
triggers as ``media reports'' (as opposed to grooming, a considerably smaller
category). During analysis, Discord's statistics stuck out: They were
considerably smaller than \V{NCMEC}'s counts and, according to Fisher's exact
test, definitely not random outliers. When we revisited the transparency report,
we noticed a boldface ``6,948 \emph{accounts} to \V{NCMEC}'' (emphasis ours) in
the paragraph below the chart. It is eventually followed by a disclaimer that
the ``figure represents distinct accounts reported and not the total number of
reports submitted.''

We briefly considered listing Discord as one of the audited firms and then
revealing the mistaken metric in the analysis, thus drawing attention to the
real-world impact of such misleading disclosures. But by doing so, we would have
recreated the conditions of Discord's disclosure, i.e., hiding critical
information in the body text. Needless to say, that would be highly
inappropriate for a scholarly article. Though it might provide a good starting
point for a classroom exercise.

Discord is not the first technology firm to replace CyberTipline report counts
with reported account numbers. Pre-Musk Twitter did so as well. It shifts the
focus from measuring (some approximation of) incidents to measuring (some
approximation of) disruptive individuals. It also reduces the magnitude of
disclosed quantities. For instance, Discord's flagged accounts are between 34\%
and 83\% of the number of reports \V{NCMEC} received from the firm. That may
make this metric more attractive to comms and marketing. But it also calls into
question the firms' often emphasized ``zero tolerance'' when it comes to the
sexual exploitation of minors. After all, if the number of accounts is smaller
than the number of incidents, some of those accounts must have caused several
incidents. Since zero tolerance implies that offending accounts are immediately
deactivated on discovery of the first incident, the question becomes how did
they cause additional incidents? Were they discovered after the fact because of
increased account scrutiny including history? Or how do you account for this
discrepancy?

We have observed more dubious behavior as well. Reddit stands out amongst the
surveyed providers because it is the only one that has been disclosing
CyberTipline report counts since 2019. Furthermore, its disclosed counts are the
same as \V{NCMEC}'s for four out of the last five years. In the past, Reddit has
committed to providing meaningful notice about its content moderation decisions
and also to abiding by the Santa Clara Principles for content
moderation\TechReport{~\cite{Reddit2022}}. Those commitments helped make it the
only technology platform to receive five stars in the Electronic Frontier
Foundation's Who Has Your Back transparency
rankings\TechReport{~\cite{CrockerGebhartea2019}}. In short, Reddit appears to
be exemplary when it comes to transparency. Except, as we discovered during the
research for this study, Reddit has been shadow moderating comments for years,
displaying removed comments only to their
authors\TechReport{~\cite{Hawkins2023}}. That, of course, is the opposite of
meaningful notice and also a direct violation of the Santa Clara Principles.

Of course, there is a simpler option available to technology firms: Skip
transparency disclosures altogether. In fact, Apple goes even further, and after
its attempt at client-side scanning for \V{CSAM}
failed~\cite{FoxBrewsterLevine2023}, the firm gave up on the issue, forgoing
even cloud-based scanning. While pervasive encryption probably would limit what
content can be scanned, Meta's WhatsApp also faces that problem and nonetheless
managed to file 1.4 million CyberTipline reports in 2023. Instead, Apple chooses
not to scan for \V{CSAM} and not to make transparency disclosures, a degree of
inaction only shared with Quora and possibly Omegle amongst surveyed
organizations. Doing so is legal---as long as Apple files CyberTipline reports
about incidents as it becomes aware of them, which it did 267 times in 2023.
Only the Wikimedia Foundation filed fewer reports.}

\subsection{Incentives for Improving Transparency}

\begin{table}
\centering\libertineLF

\caption{Summary of surveyed organizations and their transparency practices.}
\label{tab:org:summary}

\begin{tabular}{lrr}
\toprule
\multicolumn{2}{l}{\emph{Providers That Make\ldots} \hfill\emph{Count}} & \emph{Grade} \\
\midrule
Reasonably accurate disclosures     &  7 & \multirow{2}{*}{Pass} \\
Not enough disclosures (yet)        &  1 \\
\midrule
Problematic disclosures             &  3 & \multirow{2}{*}{Fail} \\
No disclosures                      &  5 \\
\midrule
\emph{Total}                        & 16 & \\
\bottomrule
\end{tabular}
\end{table}

Table~\ref{tab:org:summary} captures a final tally of organizations (other than
\V{NCMEC}) across four transparency disclosure categories. First, all seven
firms participating in the first audit make reasonably accurate disclosures.
Second, X has only released a half year's worth of transparency data and hence
could not yet be audited. Third, Discord, Meta, and TikTok have made problematic
transparency disclosures that render some or all of their data useless. Fourth,
five more organizations skip transparency disclosures about \V{CSE} altogether.
In other words, one half of surveyed providers pass their reviews. The other
half of surveyed providers fail their reviews. That may be better than we
expected, but it also does not boost confidence in the technology industry's
ability to understand let alone moderate their social impact.

One obvious remedy for providers failing their transparency reviews is a legal
mandate that compells firms to track and disclose specific metrics. Alas, when
taking the results of the 2024 \V{US} presidential election into account, that
remedy also appears unrealistic for the forseeable future. At the same time,
this study illustrates the very benefits of such a mandate, one anchored in
\V{US} law nonetheless. Before the April 2023 release of \V{NCMEC}'s
transparency report to the Congressional Appropriations
Committees\TechReport{~\cite{NcmecOjjdp2022}}, several statistics including
long-term growth in CyberTipline reports, kinds of reported incidents, number
and uniqueness of attachments, and relationship between perpetrators and victims
were not publicly available. However, thanks to an addition to the explanatory
statement for the 2023 Appropriations Bill\TechReport{~\cite{DeLauro2022}},
\V{NCMEC} was compelled to produce the corresponding transparency data. It made
a real difference for this study.

Even if that particular stick remains out of play, we strongly believe that
upping their game in transparency reporting is in the best interest of both
industry and \V{NCMEC} as well. We belabored the reasons for needing more
granular data in Sections~\ref{sec:discussion:legal}
and~\ref{sec:discussion:deluge} already. So here, we add only two more points.
For one, social media firms probably won't escape the Brussels
effect\TechReport{~\cite{Bradford2020}} and already are subject to the Digital
Services Act's transparenyc reporting
requirements\TechReport{~\cite{EuropeanParliamentAndCouncil2022}}. They might as
well do the same reporting on a global scale. For another, a lack of data
provides a fertile breeding ground for various stakeholders' biased and
simplistic accounts when the reality, as this study illustrates for \V{CSE},
often is far more nuanced and complicated. Admittedly, better data doesn't
automagically result in more nuanced story telling (whether by press, \V{NGO}s,
or academics), far from it. But it at least allows for that to happen.

For trust and safety professionals trying to make such transparency disclosures,
we have three actionable recommendations based on our experiences with curating
transparency data. First, strongly prefer absolute metrics over relative ones.
People working with the data can easily turn absolute quantities into
normalized, fractional ones, whereas the reverse transformation usually is
harder and loses precision. Second, disclose all metrics, including historical
quantities, without rounding, in \V{CSV} format with release notes. Keeping
comms or marketing out of transparency reporting may be too much to ask for. But
please limit their spin to the written report with pretty graphs. Third, close
the loop, for example, by having freshly hired data scientists write reports
based on the available data. Then incorporate their feedback.

\section{Conclusion}
\label{sec:conclusion}

In summary, the evidentiary record does \emph{not} support the more
sensationalist claims about child sexual exploitation. There is \emph{no}
``explosion'' of \V{CSAM}. The internet is \emph{not} ``overrun with images of
child sexual abuse.'' The pandemic did \emph{not} trigger more sexual
exploitation overall, even if there was a marked spike in March 2020.

Nonetheless, the evidentiary record does support the claim that \V{US}-based
social media with their unprecedented global reach and propensity for virality,
the near total legal prohibition against sexual imagery involving minors, and
perfectly normal adolescent exploration of sexuality, boundaries, relationships,
and so on have combined to create a perfect storm of CyberTipline reports. And
that storm is now threatening to overwhelm the very organization serving as a
critical global resource. Not only that, but the majority of reports is also
useless. That should suffice as encouragement for legislative reform. The fact
that laws intended to protect minors are increasingly criminalizing and thereby
harming them should add urgency. That also makes language such as
``self-generated \V{CSAM}'' counter-productive.

As far as transparency is concerned, many technology firms did far better than
we expected. As this study has amply demonstrated, transparency disclosures
needn't be mere transparency theater, but can help produce real insight. The
most surprising result from this study is the first audit's finding that firms
generally disclose reasonably accurate CyberTipline report counts but
nonetheless have a very small but statistically significant bias towards
underreporting. In other words, the professionalism of the trust and safety
personnel preparing these statistics prevails, but it cannot entirely overcome
all too human biases. Making such discoveries requires access to meaningful
data. As this study has shown, we do not have enough when it comes to online
sexual exploitation of children.

\begin{anonsuppress}
\begin{acks}
An \V{NCMEC} employee helped complicate our thinking about \V{CSAM}---while
somehow also avoiding to answer all of our questions. Meanwhile Karin Wolman did
answer our questions about 18 \V{US} Code. Jan Vitek provided helpful feedback
on earlier versions of this article. Thank you!

This work was supported in part by \V{MEYS}, \V{ERC} \V{CZ} program, grant no.\
LL2325 and by National Science Foundation grants CCF-2139612 and CCF-1910850.
\end{acks}

\end{anonsuppress}

\bibliographystyle{ACM-Reference-Format}
\bibliography{bibliography}

\TechReport{
\appendix
\section{The Data}
\label{sec:appendix:data}

\begin{table}
\centering\libertineLF

\caption{CSAM pieces and CyberTipline reports disclosed by technology firms and NCMEC: 2019 to 2023 (pt.\ 1)}
\label{tab:pieces-and-reports-1}

\begin{tabular}{r@{\hskip 1.5em}rrr r@{\hskip 1.5em}rrr}
\toprule
& \multicolumn{3}{c}{Disclosed by Service Provider}
& \hspace{3.5em}
& \multicolumn{3}{c}{Disclosed by NCMEC} \\
\cmidrule{2-4}\cmidrule{6-8}

Year
& \multicolumn{1}{c}{Pieces}
& \multicolumn{1}{c}{per}
& \multicolumn{1}{c}{Reports}
& \multicolumn{1}{c}{$\Delta$\%}
& \multicolumn{1}{c}{Reports}
& \multicolumn{1}{c}{of (\%)}
& \multicolumn{1}{c}{Total}\\[2ex]

& \multicolumn{3}{c}{\textsc{\MakeLowercase{Meta (Q\;\ImageCheck\VideoCheck)}}} & & & & \\ \cmidrule{2-4}
\color{lowlight} 2019 &  39,368,400 &   2.48 &            &         & 15,884,511 & 93.508 & \color{lowlight} 16,987,361 \\
\color{lowlight} 2020 &  38,890,800 &   1.92 &            &         & 20,307,216 & 93.362 & \color{lowlight} 21,751,085 \\
\color{lowlight} 2021 &  78,012,400 &   2.90 &            &         & 26,885,302 & 91.454 & \color{lowlight} 29,397,681 \\
\color{lowlight} 2022 & 105,800,000 &   3.89 &            &         & 27,190,665 & 84.814 & \color{lowlight} 32,059,029 \\
\color{lowlight} 2023 &  63,300,000 &   2.07 &            &         & 30,658,047 & 84.666 & \color{lowlight} 36,210,368 \\
\addlinespace

& \multicolumn{3}{c}{\textsc{\MakeLowercase{Google (H\;\ImageCheck\VideoCheck)}}} & & & & \\ \cmidrule{2-4}
\color{lowlight} 2019 &             &        &            &         &    449,283 &  2.645 & \color{lowlight} 16,987,361 \\
\color{lowlight} 2020 &   4,437,853 &   8.10 &    547,875 &   -0.21 &    546,704 &  2.513 & \color{lowlight} 21,751,085 \\
\color{lowlight} 2021 &   6,696,497 &   7.69 &    870,319 &   +0.63 &    875,783 &  2.979 & \color{lowlight} 29,397,681 \\
\color{lowlight} 2022 &  13,402,885 &   6.16 &  2,174,319 &   +0.01 &  2,174,548 &  6.783 & \color{lowlight} 32,059,029 \\
\color{lowlight} 2023 &   7,955,169 &   5.40 &  1,472,221 &   -0.09 &  1,470,958 &  4.062 & \color{lowlight} 36,210,368 \\
\addlinespace

& \multicolumn{3}{c}{\textsc{\MakeLowercase{X née Twitter (H\;\ImageCheck\VideoCheck)}}} & & & & \\ \cmidrule{2-4}
\color{lowlight} 2019 &             &        &           &          &     45,726 &  0.269 & \color{lowlight} 16,987,361 \\
\color{lowlight} 2020 &             &        &           &          &     65,062 &  0.299 & \color{lowlight} 21,751,085 \\
\color{lowlight} 2021 &             &        &           &          &     86,666 &  0.295 & \color{lowlight} 29,397,681 \\
\color{lowlight} 2022 &             &        &           &          &     98,050 &  0.306 & \color{lowlight} 32,059,029 \\
\color{lowlight} 2023 &             &        &           &          &    870,503 &  2.404 & \color{lowlight} 36,210,368 \\
\addlinespace

& \multicolumn{3}{c}{\textsc{\MakeLowercase{Snap (H\;\ImageCheck\VideoCheck)}}} & & & & \\ \cmidrule{2-4}
\color{lowlight} 2019 &             &        &            &         &     82,030 &  0.483 & \color{lowlight} 16,987,361 \\
\color{lowlight} 2020 &             &        &            &         &    144,095 &  0.662 & \color{lowlight} 21,751,085 \\
\color{lowlight} 2021 &             &        &            &         &    512,522 &  1.743 & \color{lowlight} 29,397,681 \\
\color{lowlight} 2022 &   1,273,838 &   2.31 &    550,755 &   +0.06 &    551,086 &  1.719 & \color{lowlight} 32,059,029 \\
\color{lowlight} 2023 &   1,594,805 &   2.31 &    691,225 &   +3.11 &    713,055 &  1.969 & \color{lowlight} 36,210,368 \\
\addlinespace

& \multicolumn{3}{c}{\textsc{\MakeLowercase{TikTok (Q\;\ImageCheck\VideoCheck)}}} & & & & \\ \cmidrule{2-4}
\color{lowlight} 2019 &             &        &           &          &        596 &  0.004 & \color{lowlight} 16,987,361 \\
\color{lowlight} 2020 &             &        &           &          &     22,692 &  0.104 & \color{lowlight} 21,751,085 \\
\color{lowlight} 2021 &             &        &           &          &    154,618 &  0.526 & \color{lowlight} 29,397,681 \\
\color{lowlight} 2022 &             &        &           &          &    288,125 &  0.899 & \color{lowlight} 32,059,029 \\
\color{lowlight} 2023 &             &        &           &          &    590,376 &  1.630 & \color{lowlight} 36,210,368 \\
\addlinespace

& \multicolumn{3}{c}{\textsc{\MakeLowercase{Discord (Q\;\ImageCheck)}}} & & & & \\ \cmidrule{2-4}
\color{lowlight} 2019 &             &        &           &          &     19,480 &  0.115 & \color{lowlight} 16,987,361 \\
\color{lowlight} 2020 &             &        &           &          &     15,324 &  0.071 & \color{lowlight} 21,751,085 \\
\color{lowlight} 2021 &             &        &           &          &     29,606 &  0.101 & \color{lowlight} 29,397,681 \\
\color{lowlight} 2022 &             &        &           &          &    169,800 &  0.530 & \color{lowlight} 32,059,029 \\
\color{lowlight} 2023 &             &        &           &          &    339,412 &  0.937 & \color{lowlight} 36,210,368 \\
\addlinespace

& \multicolumn{3}{c}{\textsc{\MakeLowercase{Reddit (H\;\ImageCheck\VideoCheck)}}} & & & & \\ \cmidrule{2-4}
\color{lowlight} 2019 &             &        &        724 &   \same &        724 &  0.004 & \color{lowlight} 16,987,361 \\
\color{lowlight} 2020 &             &        &      2,233 &   \same &      2,233 &  0.010 & \color{lowlight} 21,751,085 \\
\color{lowlight} 2021 &       9,258 &   0.92 &     10,059 &   \same &     10,059 &  0.034 & \color{lowlight} 29,397,681 \\
\color{lowlight} 2022 &      80,888 &   1.54 &     52,592 &   \same &     52,592 &  0.164 & \color{lowlight} 32,059,029 \\
\color{lowlight} 2023 &             &        &    290,121 &   +0.01 &    290,141 &  0.801 & \color{lowlight} 36,210,368 \\

\bottomrule
\end{tabular}
\end{table}

\begin{table}
\centering\libertineLF

\caption{CSAM pieces and CyberTipline reports disclosed by technology firms and NCMEC: 2019 to 2023 (pt.\ 2)}
\label{tab:pieces-and-reports-2}

\begin{tabular}{r@{\hskip 1.5em}rrr r@{\hskip 1.5em}rrr}
\toprule
& \multicolumn{3}{c}{Disclosed by Service Provider}
& \hspace{3.5em}
& \multicolumn{3}{c}{Disclosed by NCMEC} \\
\cmidrule{2-4}\cmidrule{6-8}

Year
& \multicolumn{1}{c}{Pieces}
& \multicolumn{1}{c}{per}
& \multicolumn{1}{c}{Reports}
& \multicolumn{1}{c}{$\Delta$\%}
& \multicolumn{1}{c}{Reports}
& \multicolumn{1}{c}{of (\%)}
& \multicolumn{1}{c}{Total}\\[2ex]

& \multicolumn{3}{c}{\textsc{\MakeLowercase{Omegle}}} & & & & \\ \cmidrule{2-4}
\color{lowlight} 2019 &             &        &           &          &      3,470 &  0.020 & \color{lowlight} 16,987,361 \\
\color{lowlight} 2020 &             &        &           &          &     20,265 &  0.093 & \color{lowlight} 21,751,085 \\
\color{lowlight} 2021 &             &        &           &          &     46,924 &  0.160 & \color{lowlight} 29,397,681 \\
\color{lowlight} 2022 &             &        &           &          &    608,601 &  1.898 & \color{lowlight} 32,059,029 \\
\color{lowlight} 2023 &             &        &           &          &    188,102 &  0.520 & \color{lowlight} 36,210,368 \\
\addlinespace

& \multicolumn{3}{c}{\textsc{\MakeLowercase{Microsoft (H\;\ImageCheck\VideoCheck)}}} & & & & \\ \cmidrule{2-4}
\color{lowlight} 2019 &             &        &            &         &    123,927 &  0.730 & \color{lowlight} 16,987,361 \\
\color{lowlight} 2020 &   1,256,652 &  13.03 &     96,435 &   +0.42 &     96,836 &  0.445 & \color{lowlight} 21,751,085 \\
\color{lowlight} 2021 &     564,383 &   7.12 &     78,926 &   -0.05 &     78,883 &  0.268 & \color{lowlight} 29,397,681 \\
\color{lowlight} 2022 &     452,384 &   4.22 &    107,599 &   +1.11 &    108,798 &  0.339 & \color{lowlight} 32,059,029 \\
\color{lowlight} 2023 &     402,630 &   2.86 &    140,720 &   +0.37 &    141,236 &  0.390 & \color{lowlight} 36,210,368 \\
\addlinespace

& \multicolumn{3}{c}{\textsc{\MakeLowercase{Pinterest (Q/H\;\ImageCheck\VideoCheck)}}} & & & & \\ \cmidrule{2-4}
\color{lowlight} 2019 &             &        &            &         &      7,360 &  0.043 & \color{lowlight} 16,987,361 \\
\color{lowlight} 2020 &             &        &      3,432 &   \same &      3,432 &  0.016 & \color{lowlight} 21,751,085 \\
\color{lowlight} 2021 &       1,608 &   0.60 &      2,684 &  -16.15 &      2,283 &  0.008 & \color{lowlight} 29,397,681 \\
\color{lowlight} 2022 &      37,136 &   1.13 &     32,964 &   +4.00 &     34,310 &  0.107 & \color{lowlight} 32,059,029 \\
\color{lowlight} 2023 &      57,774 &   1.15 &     50,437 &   +3.73 &     52,356 &  0.145 & \color{lowlight} 36,210,368 \\
\addlinespace

& \multicolumn{3}{c}{\textsc{\MakeLowercase{Amazon (Y\;\ImageCheck)}}} & & & & \\ \cmidrule{2-4}
\color{lowlight} 2019 &             &        &           &          &        549 &  0.003 & \color{lowlight} 16,987,361 \\
\color{lowlight} 2020 &             &        &     2,235 &    \same &      2,235 &  0.010 & \color{lowlight} 21,751,085 \\
\color{lowlight} 2021 &      27,244 &   0.81 &    33,848 &    -0.04 &     33,833 &  0.116 & \color{lowlight} 29,397,681 \\
\color{lowlight} 2022 &      52,656 &   0.79 &    67,073 &    +4.49 &     70,157 &  0.220 & \color{lowlight} 32,059,029 \\
\color{lowlight} 2023 &      24,756 &   0.79 &    31,281 &    +3.39 &     32,359 &  0.090 & \color{lowlight} 36,210,368 \\
\addlinespace

& \multicolumn{3}{c}{\textsc{\MakeLowercase{Automattic (\ImageCheck)}}} & & & & \\ \cmidrule{2-4}
\color{lowlight} 2019 &             &        &           &          &     10,443 &  0.062 & \color{lowlight} 16,987,361 \\
\color{lowlight} 2020 &             &        &           &          &      9,130 &  0.042 & \color{lowlight} 21,751,085 \\
\color{lowlight} 2021 &             &        &           &          &      4,821 &  0.016 & \color{lowlight} 29,397,681 \\
\color{lowlight} 2022 &             &        &           &          &      5,035 &  0.016 & \color{lowlight} 32,059,029 \\
\color{lowlight} 2023 &             &        &           &          &     19,591 &  0.054 & \color{lowlight} 36,210,368 \\

& \multicolumn{3}{c}{\textsc{\MakeLowercase{Quora}}} & & & & \\ \cmidrule{2-4}
\color{lowlight} 2019 &             &        &           &          &          1 & $\qtiny$ & \color{lowlight} 16,987,361 \\
\color{lowlight} 2020 &             &        &           &          &          2 & $\qtiny$ & \color{lowlight} 21,751,085 \\
\color{lowlight} 2021 &             &        &           &          &         25 & $\qtiny$ & \color{lowlight} 29,397,681 \\
\color{lowlight} 2022 &             &        &           &          &      2,242 &    0.007 & \color{lowlight} 32,059,029 \\
\color{lowlight} 2023 &             &        &           &          &      6,135 &    0.017 & \color{lowlight} 36,210,368 \\
\addlinespace

& \multicolumn{3}{c}{\textsc{\MakeLowercase{Aylo née MindGeek (H\;\ImageCheck\VideoCheck)}}} & & & & \\ \cmidrule{2-4}
\color{lowlight} 2019 &             &        &           &          &            &        & \color{lowlight} 16,987,361 \\
\color{lowlight} 2020 &             &        &     4,171 &  +104.12 &     13,229 &  0.061 & \color{lowlight} 21,751,085 \\
\color{lowlight} 2021 &      20,401 &   2.26 &     9,029 &    +0.82 &      9,103 &  0.031 & \color{lowlight} 29,397,681 \\
\color{lowlight} 2022 &       9,588 &   4.80 &     1,996 &    +4.84 &      2,095 &  0.007 & \color{lowlight} 32,059,029 \\
\color{lowlight} 2023 &       7,313 &   2.92 &     2,503 &    +3.65 &      2,596 &  0.007 & \color{lowlight} 36,210,368 \\
\addlinespace

\bottomrule
\end{tabular}
\end{table}

\begin{table}
\centering\libertineLF

\caption{CSAM pieces and CyberTipline reports disclosed by technology firms and NCMEC: 2019 to 2023 (pt.\ 3)}
\label{tab:pieces-and-reports-3}

\begin{tabular}{r@{\hskip 1.5em}rrr r@{\hskip 1.5em}rrr}
\toprule
& \multicolumn{3}{c}{Disclosed by Service Provider}
& \hspace{3.5em}
& \multicolumn{3}{c}{Disclosed by NCMEC} \\
\cmidrule{2-4}\cmidrule{6-8}

Year
& \multicolumn{1}{c}{Pieces}
& \multicolumn{1}{c}{per}
& \multicolumn{1}{c}{Reports}
& \multicolumn{1}{c}{$\Delta$\%}
& \multicolumn{1}{c}{Reports}
& \multicolumn{1}{c}{of (\%)}
& \multicolumn{1}{c}{Total}\\[2ex]

& \multicolumn{3}{c}{\textsc{\MakeLowercase{Apple}}} & & & & \\ \cmidrule{2-4}
\color{lowlight} 2019 &             &        &           &          &        205 & $\qtiny$ & \color{lowlight} 16,987,361 \\
\color{lowlight} 2020 &             &        &           &          &        265 & $\qtiny$ & \color{lowlight} 21,751,085 \\
\color{lowlight} 2021 &             &        &           &          &        160 & $\qtiny$ & \color{lowlight} 29,397,681 \\
\color{lowlight} 2022 &             &        &           &          &        234 & $\qtiny$ & \color{lowlight} 32,059,029 \\
\color{lowlight} 2023 &             &        &           &          &        267 & $\qtiny$ & \color{lowlight} 36,210,368 \\
\addlinespace

& \multicolumn{3}{c}{\textsc{\MakeLowercase{Wikimedia (\ImageCheck)}}} & & & & \\ \cmidrule{2-4}
\color{lowlight} 2019 &             &        &           &          &         13 & $\qtiny$ & \color{lowlight} 16,987,361 \\
\color{lowlight} 2020 &             &        &           &          &         11 & $\qtiny$ & \color{lowlight} 21,751,085 \\
\color{lowlight} 2021 &             &        &           &          &          8 & $\qtiny$ & \color{lowlight} 29,397,681 \\
\color{lowlight} 2022 &             &        &           &          &         29 & $\qtiny$ & \color{lowlight} 32,059,029 \\
\color{lowlight} 2023 &             &        &           &          &         34 & $\qtiny$ & \color{lowlight} 36,210,368 \\

\bottomrule
\end{tabular}
\end{table}

Tables~\ref{tab:pieces-and-reports-1}--\ref{tab:pieces-and-reports-3} provide
the data released by electronic service providers and \V{NCMEC} about
CyberTipline reports and attached pieces, aggregated by corporation and year.
Parenthesized material after provider names repeats the disclosure frequencies
from Table~\ref{tab:survey} as Q/H/Y for every 4/6/12 months, respectively, and
then adds picture and/or video icons if the provider proactively scans for
\V{CSAM}. Pinterest is marked ``Q/H'' because the firm releases quarterly
aggregates every six months.

The $\Delta\%$ column shows the percentage difference between the two report
counts in terms of the mean, just as in Figure~\ref{fig:reports-audit}. Some of
the entries are, in fact, identical; the corresponding entries are marked as
$\equiv$. Also, some of the percentages for Quora, Apple, and the Wikimedia
Foundation in the middle column for \V{NCMEC}'s data are too small to fit into
the alloted three-digit precision and hence are marked as $\qtiny$.

Since years in the first column and total report counts in the last column are
repeated for every service provider, they are presented in gray to make tables
look less busy. The open space for providers that do not make transparency
disclosures strengthens that effect. Alas, the motivation for giving every
provider the same table real estate was to avoid biasing the presentation.

Providers are sorted in descending order of report counts for 2023.
 }

\end{document}